\documentclass[twocolumn]{aastex631}

\usepackage{graphicx} 
\usepackage{booktabs} 
\usepackage{verbatim}
\usepackage{csquotes}
\usepackage{textcmds}
\usepackage{footmisc}
\usepackage{float}

\begin{document}

\title{Machine Learning for Radial Velocity Analysis I: Vision Transformers as a Robust Alternative for Detecting Planetary Candidates}

\author[0009-0004-9702-5243]{Anoop Gavankar}
\affiliation{Department of Astronomy and Astrophysics, Tata Institute of Fundamental Research, Homi Bhabha Road, Colaba, Mumbai 400005, India}
\email{anoop.gavankar@tifr.res.in}

\author{Tanish Mittal}
\affiliation{Birla Institute of Technology, Pilani, Rajasthan 333031, India}
\email{tanishmittal0658@gmail.com}


\author[0000-0001-8720-5612]{Joe P.\ Ninan}
\affiliation{Department of Astronomy and Astrophysics, Tata Institute of Fundamental Research, Homi Bhabha Road, Colaba, Mumbai 400005, India}
\email{indiajoe@gmail.com}

\author[0000-0003-2896-1471]{Shravan Hanasoge}
\affiliation{Department of Astronomy and Astrophysics, Tata Institute of Fundamental Research, Homi Bhabha Road, Colaba, Mumbai 400005, India}
\email{hanasoge@tifr.res.in}

\begin{abstract}

Extreme precision radial velocity (EPRV) surveys usually require extensive observational baselines to confirm planetary candidates, making them resource-intensive. Traditionally, periodograms are used to identify promising candidate signals before further observational investment, but their effectiveness is often limited for low-amplitude signals due to stellar jitter. In this work, we develop a machine learning (ML) framework based on a Transformer architecture that aims to detect the presence and likely period of planetary signals in time-series spectra, even in the presence of stellar activity. The model is trained to classify whether a planetary signal exists and assign it to one of several discrete period and amplitude bins. Injection-recovery tests on randomly selected 100 epoch observation subsets from NEID solar data (2020-2022 period) show that for low-amplitude systems ($<$1 ms$^{-1}$), our model improves planetary candidate identification by a factor of two compared to the traditional Lomb-Scargle periodogram.

Our ML model is built on a Vision Transformer (ViT) architecture that processes reduced representations of solar spectrum observations to predict the period and semi-amplitude of planetary signal candidates. By analyzing multi-epoch spectra, the model reliably detects planetary signals with semi-amplitudes as low as 65 cms$^{-1}$. Even under real solar noise and irregular sampling, it identifies signals down to 35 cms$^{-1}$. Comparisons with the Lomb-Scargle periodogram demonstrate a significant improvement in detecting low-amplitude planetary candidates, particularly for longer orbital periods. These results underscore the potential of machine learning to identify planetary candidates early in EPRV surveys, even from limited observational counts.

\end{abstract}

\section{Introduction} \label{sec:intro}

The discovery and characterization of exoplanets have become central to modern astrophysics, offering key insights into planetary formation and evolution. One of the most widely used techniques for detecting these distant worlds is the radial velocity (RV) method. Since the first exoplanet detection via RV measurements \citep{1995Natur.378..355M}, Doppler reflex observations have advanced significantly, enabling the precise characterization of planetary systems beyond the Solar System.

The RV method detects exoplanets by measuring Doppler shifts in a star’s spectral lines caused by the gravitational pull of an orbiting planet. However, these measurements are affected by surface phenomena on the host star, collectively referred to as stellar jitter, which introduces noise and complicates planet detection.

In an Earth-Sun system, the RV semi-amplitude is about 9 cms$^{-1}$. However, stellar RV measurements are typically precise to within approximately 1 ms$^{-1}$ \citep{2016MNRAS.457.3637H, Dumusque_2018}, primarily due to the limiting effects of stellar jitter. Thus, detecting Earth-mass exoplanets in the habitable zones of Sun-like stars requires improving our RV measurement error margin by an order of magnitude.

Current high-resolution spectrographs such as HARPS-N (High Accuracy Radial velocity Planet Searcher for the Northern hemisphere)\citep{2012SPIE.8446E..1VC}, ESPRESSO (Echelle SPectrograph for Rocky Exoplanet and Stable Spectroscopic Observations)  \citep{2021A&A...645A..96P}, CARMENES (Calar Alto high-Resolution search for M dwarfs with Exoearths with Near-infrared and optical Échelle Spectrograph) \citep{2018SPIE10702E..0WQ}, HPF (Habitable-Zone Planet Finder) \citep{2012SPIE.8446E..1SM}, NEID (NN-explore Exoplanet Investigations with Doppler spectroscopy) \citep{Schwab_2016SPIE.9908E..7HS, Halverson2016_10.1117/12.2232761, Paul_2019_10.1117/1.JATIS.5.1.015003}, among others, have led efforts to improve instrumental RV precision for stellar spectra. Future high-resolution spectrographs are expected to achieve the long-term RV stability necessary for detecting Earth-mass exoplanets in the habitable zones of Sun-like stars \citep{2020AJ....159..238B}.

Traditional astrophysical insight-driven methods for mitigating stellar jitter have primarily focused on targeting specific underlying sources. For example, \citet{Chaplin_2019} demonstrated that optimizing exposure times based on stellar parameters, particularly the solar-like oscillation frequency ($\nu_{\max} \approx 3.1$ mHz for the Sun), can effectively average out p-mode oscillations, reducing their impact on radial velocity measurements to within 10 cm s$^{-1}$.

Additional data-driven techniques for mitigating stellar RV activity include time-correlated modeling approaches, typically via Gaussian process modeling \citep[e.g.,][]{Haywood_2014, Rajpaul_2015, jones2020improvingexoplanetdetectionpower,Stephan_2023}, which capture correlated noise in RV datasets. Activity indicators, including H$\alpha$ \citep[][]{2007A&A...474..293B, Robertson_2014, 2014A&A...566A..35S, 10.1093/mnras/stz1215}, log $R_{HK}$ \citep{1984ApJ...279..763N}, and the Bisector Inverse Slope Span (BIS) \citep{2001A&A...379..279Q}, have also been employed to track and decorrelate activity-induced RV shifts. An alternative approach involves identifying and selectively utilizing spectral lines based on their sensitivity to stellar jitter, enabling decorrelation of activity-driven variations from planetary signals \citep{Dumusque_2018, Cretignier_2021, Wise_2022}.

\citet{Davis_2017} applied principal component analysis (PCA) to spectral data, while \citet{2022_Cretignier} utilized it on shell representation of spectra to disentangle stellar activity-induced RV variations from Keplerian motion. 

The autocorrelation function (ACF) of the cross-correlation function (CCF) is another tool used to analyze stellar jitter-related RV variations \citep{2021MNRAS.505.1699C}. Since the ACF remains invariant under Keplerian shifts, its variation is sensitive to stellar jitter, providing a direct correlation with activity-induced noise. 

Several studies have combined spectroscopic and photometric observations to mitigate stellar activity in RV measurements. The FF$^{\prime}$ method \citep{2012MNRAS.419.3147A} models activity-induced RV variations based on flux changes but relies on high-cadence observations, which are often challenging to obtain. Gaussian Process modeling extends this approach by capturing correlated noise across spectroscopic and photometric datasets \citep{Rajpaul_2015}. Disentangling techniques have also been employed to separate the impact of stellar surface features on RV variations \citep{2021ApJ...920...21M}. Additionally, decorrelating RV measurements from periodic signals linked to stellar rotation helps suppress activity-induced noise \citep{2020AJ....159..271K}. 

These methods, however, often leave valuable spectral information unutilized by relying on averaging, broad statistical representations, or selectively utilizing spectral data. Machine learning (ML) offers a promising alternative by detecting subtle deviations in spectral line configurations, potentially capturing information overlooked by traditional methods. Although this approach requires a large training dataset, it is less reliant on high-cadence observations, making it well-suited for practical observational constraints.

Neural networks have been widely applied in exoplanet research for various tasks, including exoplanet detection via the transit method \citep{10.1093/mnras/sty3146, Malik_2021, hansen2024singletransitdetectionkepler}, analysis of simulated datasets \citep{Zucker_2018, 2018MNRAS.474..478P}, and studies combining synthetic and real data \citep{Cuellar2022}. They have also been employed to distinguish planetary candidates from false positives in datasets from Kepler \citep{2018ApJ...869L...7A, 2018AJ....155...94S}, K2 \citep{2019AJ....157..169D}, TESS \citep{2019AJ....158...25Y, 2020A&A...633A..53O}, NGTS \citep{2019MNRAS.488.5232C}, and WASP \citep{2019MNRAS.483.5534S}. Additionally, these methods have been tested on confirmed Kepler exoplanets, focusing on classification and result verification \citep{Cui_2022}.

Neural networks have also found diverse applications in the RV method, addressing key challenges such as correcting radial velocities using physical observables \citep{Perger_2023}, detecting and identifying planetary signals in RV data \citep{Nieto_2023} and mitigating stellar activity signals in both simulated and solar datasets \citep{de_Beurs_2022}. In particular, Convolutional Neural Networks have been shown to enhance sensitivity to low-amplitude radial velocity signals, achieving a threshold of 0.2 ms$^{-1}$ on the HARPS-N solar dataset \citep{zhao2024improvingearthlikeplanetdetection}.

Here we introduce a Transformer‑based detection pipeline that (i) classifies whether a planetary signal is present in time-series spectra affected by stellar activity, and (ii) predicts the most likely period bin when such a signal exists. We train the model using synthetic Keplerian signals injected into NEID solar spectra \citep{Lin_2022}, enabling it to distinguish between activity-induced and planetary RV variations. Our results demonstrate that machine learning approaches can identify low-amplitude ($<$1 ms$^{-1}$) planetary signals from relatively few observations, offering improved sensitivity where traditional periodogram methods face limitations. While the model does not aim to  explicitly disentangle stellar activity from planetary signals at individual epochs of observation, it enables reliable detection and period estimation from the complete time-series data in the presence of stellar variability.

In Section \ref{sec:data}, we provide an overview of the observational data utilized in the analysis. Section \ref{sec:3} details the preprocessing steps required to prepare the data for ML algorithms. Section \ref{sec:new4} outlines the methodologies for data generation, and Section \ref{sec:4} details the architecture of our ML models and describes the training procedures implemented. The results of our investigation are presented in Section \ref{sec:5}, followed by a discussion of the implications in Section \ref{sec:6} and conclusions in Section \ref{sec:7}.

\section{Data} 
\label{sec:data}

Machine learning models require well-structured datasets for training and validation. In this study, we use high-resolution, publicly available solar observations from the NEID instrument to develop and evaluate our models.

\subsection{NEID Spectrograph}

\label{subsec:2.1}

NEID is a high-precision spectrograph designed for Doppler observations of nearby stars, installed on the 3.5-meter WIYN Telescope at Kitt Peak National Observatory. At night, it observes stellar targets, while during the day, a dedicated solar feed enables \enquote{Sun-as-a-star} measurements. With a spectral resolution of approximately 117,000, NEID delivers precise RV measurements, making it a valuable tool for exoplanetary studies \citep{Schwab_2016SPIE.9908E..7HS, Halverson2016_10.1117/12.2232761, Paul_2019_10.1117/1.JATIS.5.1.015003}.

The NEID dataset analyzed in this study consists of 19 months of solar observations from December 2020 to June 2022. The spectrograph spans a wavelength range of 380–930 nm across 122 echelle orders. Daytime observations were conducted via the NEID solar feed, with an integration time of 93 seconds per exposure \citep{Lin_2022}.

The NEID solar feed's light is attenuated to match the signal-to-noise ratio of typical NEID stellar observations. The data are processed by the NEID Data Reduction Pipeline\footnote{\label{neidpipeline}\href{https://neid.ipac.caltech.edu/docs/NEID-DRP/}{NEID data reduction pipeline}}, which converts raw spectrographic data into wavelength-calibrated solar spectra. Further details on the instrument and observational setup are available in \citet{Lin_2022}.

\begin{figure*}[ht!]
    \centering
    \includegraphics{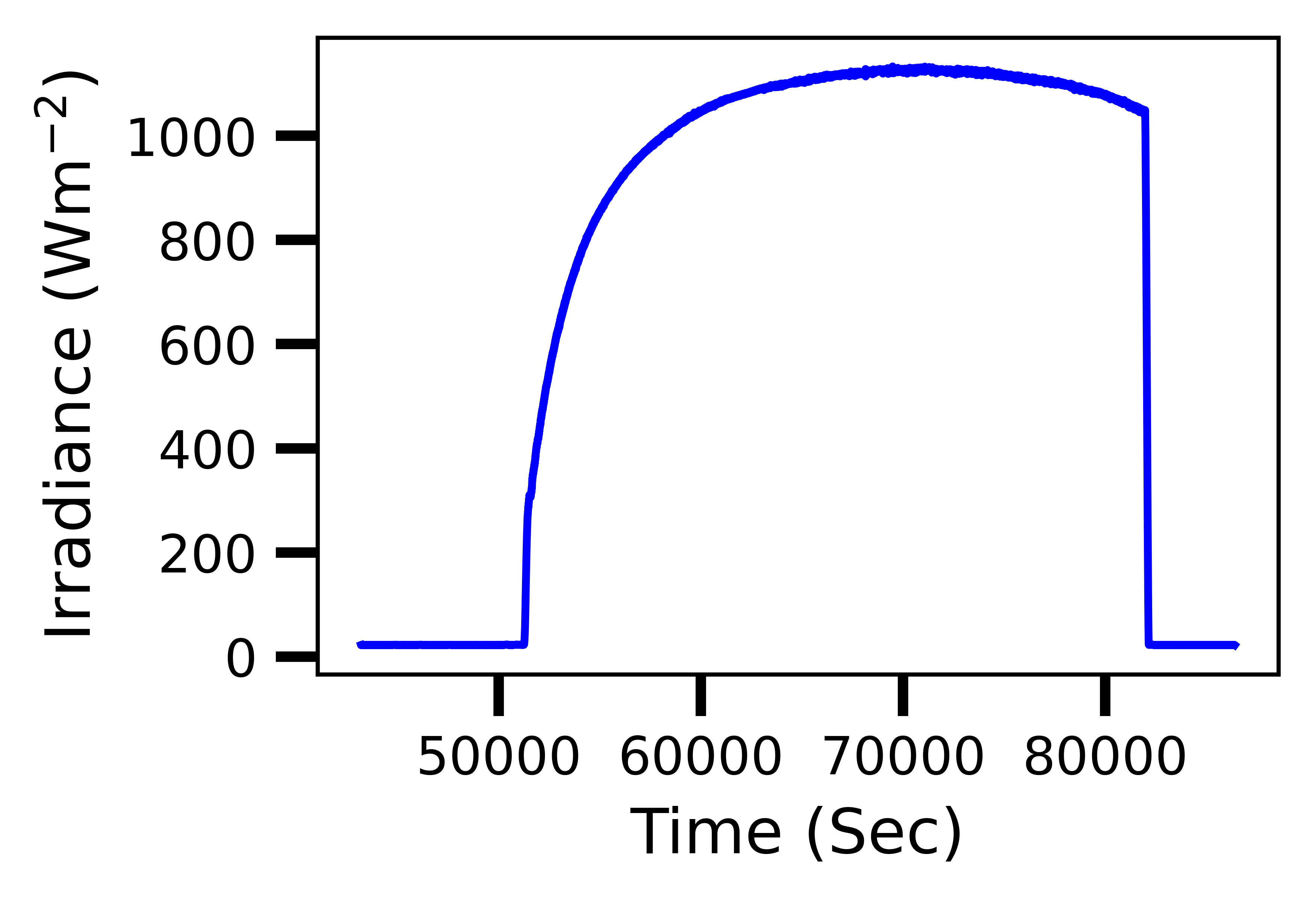}
    \hfill
    \includegraphics{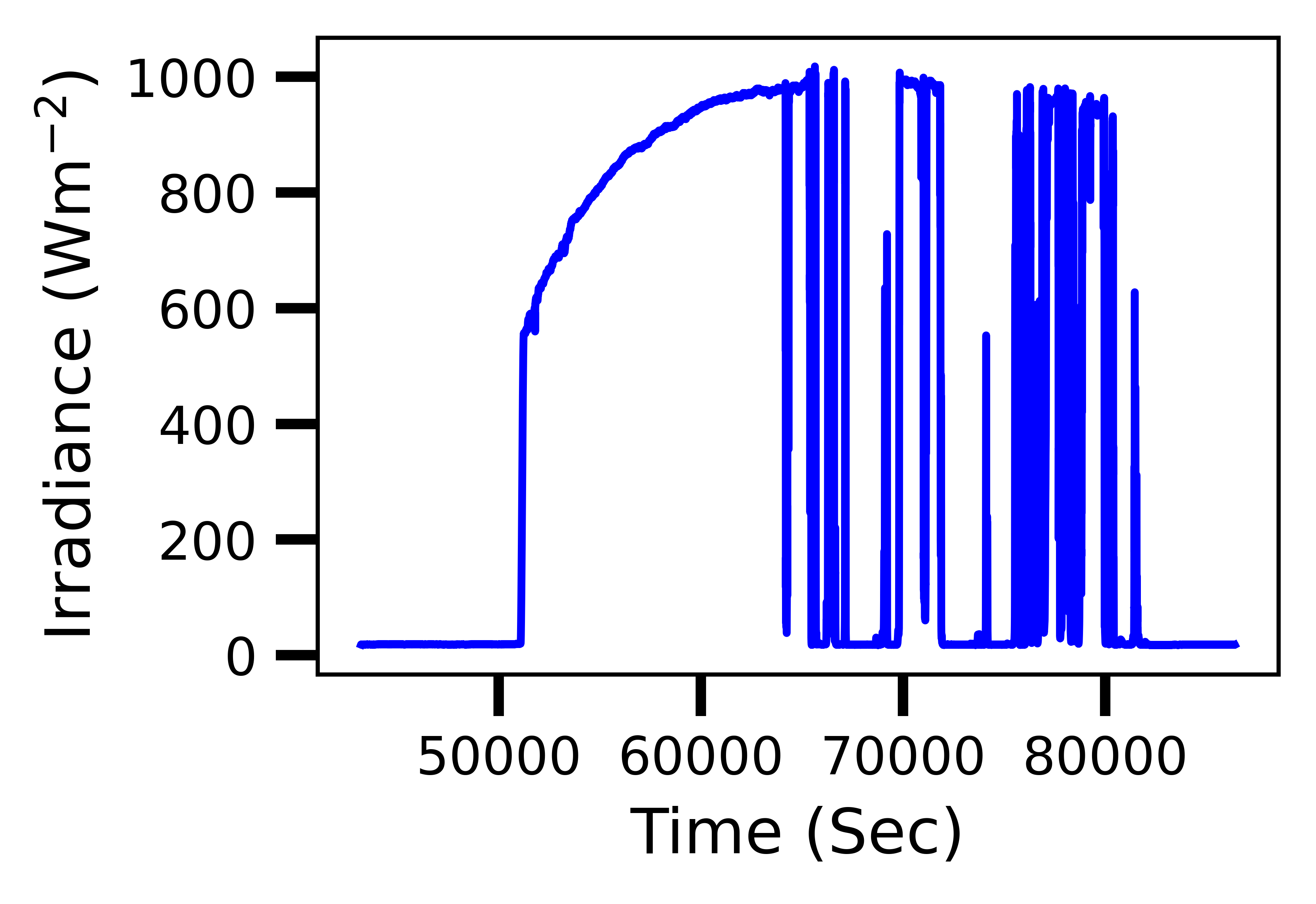}
    \caption{(a) Figure a shows the irradiance profile for a typical clear-sky day, showing smooth temporal variation with sharp transitions at dawn and dusk. The sudden flux drop at dusk is due to the shadow of the telescope building. 
    (b) Figure b shows the irradiance profile for a cloudy day, exhibiting pronounced fluctuations in solar radiation due to varying atmospheric conditions.}
    \label{fig:my_label1}
\end{figure*}

\begin{figure*}[ht!]
    \centering
    \includegraphics{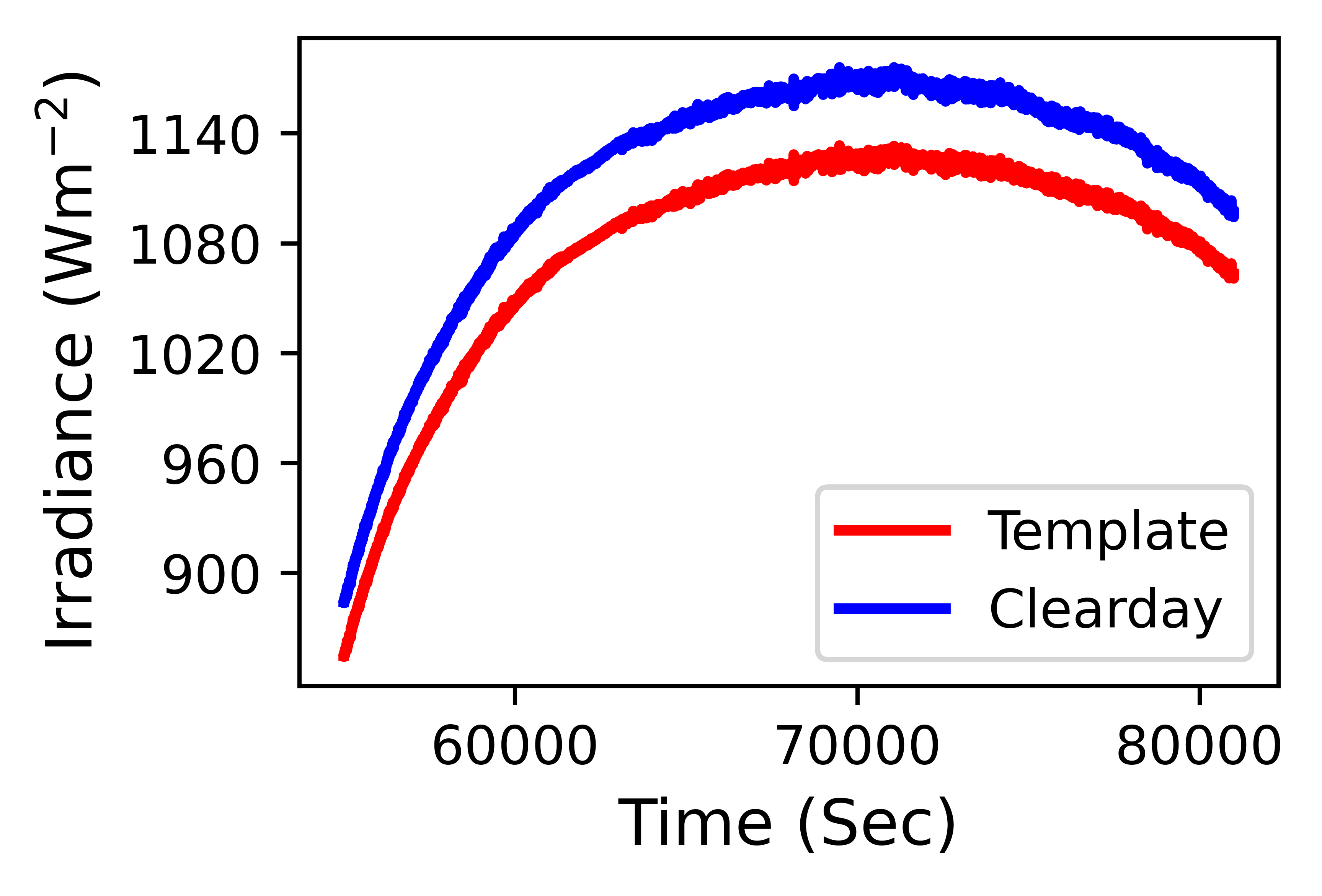}
    \hfill
    \includegraphics{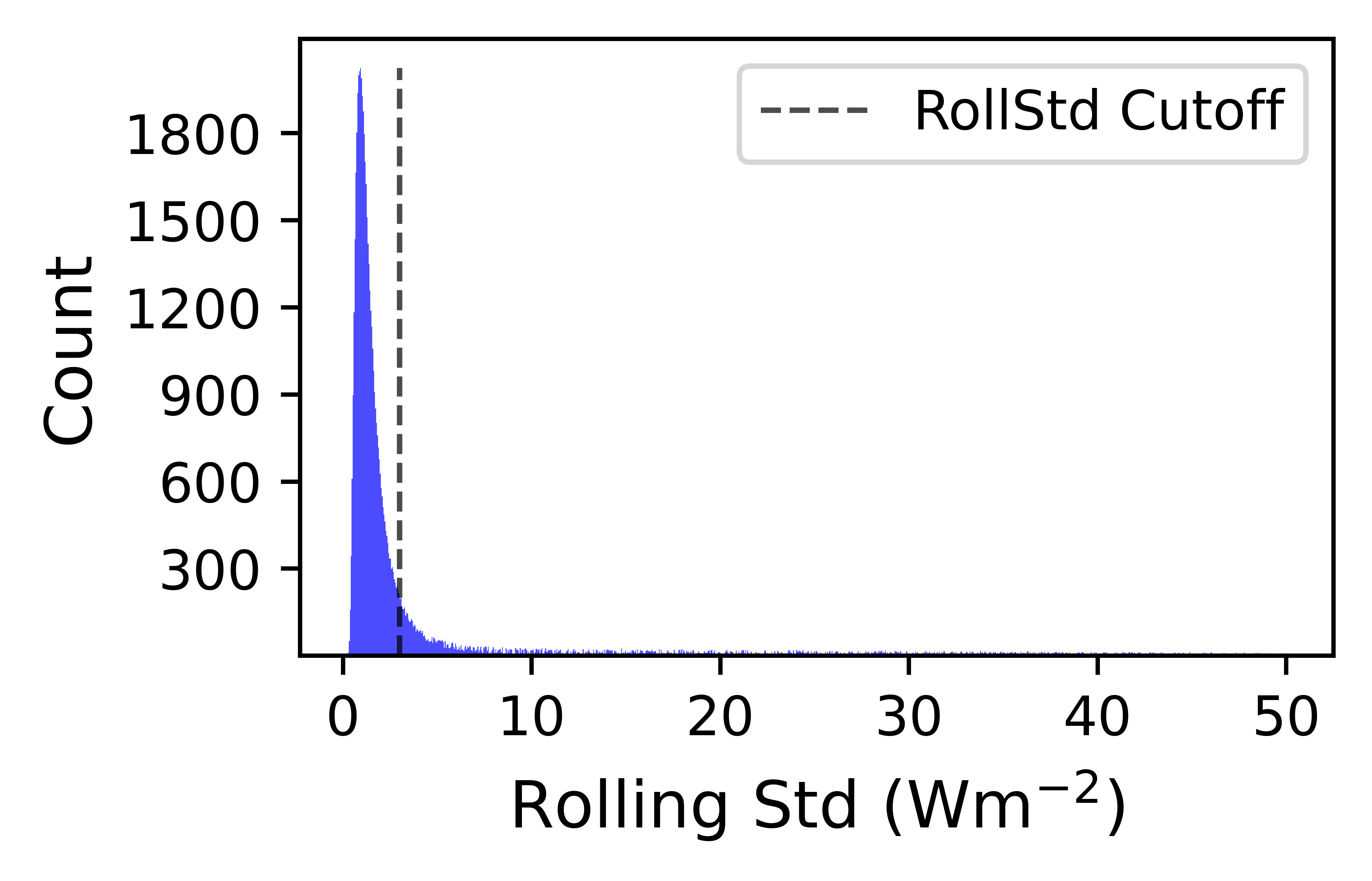}
    \caption{(a) Figure a shows the irradiance profile for a clear day compared with its monthly template from November, showing similar characteristics.  
    (b) Figure b shows a histogram of the rolling standard deviation for 30,000 randomly selected FITS file windows, displaying a pseudo-Gaussian distribution with a pronounced long tail. The chosen clear-sky day cutoff at 3 $Wm^{-2}$ is marked by the vertical dashed line.
    }
    \label{fig:my_lb2}
\end{figure*}

\section{DATA PRE-PROCESSING} 
\label{sec:3}

The NEID solar archive provides unfiltered data, encompassing all recorded solar exposures recorded by the instrument. However, the archival data cannot be directly used for training AI models. The extracted data must undergo a series of filtering, processing, and standardization steps to ensure consistency and high data quality. The procedures detailed in the following sections transform the dataset into a structured format suitable for machine learning applications.

\subsection{Filtering out Data} 

\label{subsec:3.1}

\subsubsection{Selecting Clear-Sky Data}
\label{subsec:3.1.1}

Unlike other stars, the Sun is a spatially resolved object, making its spectral lines susceptible to distortions from passing clouds that obscure different regions of the solar disk. To mitigate this effect, solar data are filtered to exclude observations affected by cloud cover.

Clear-sky periods are identified using a Pyrheliometer\footnote{\href{https://neid.ipac.caltech.edu/pyrheliometer.php}{NEID Pyrheliometer Data}} adjacent to the solar feed \citep{Lin_2022}. The pyrheliometer measurements of solar radiation intensity (Wm$^{-2}$) serve as a reference for selecting timestamps corresponding to clear observing conditions. Figure \ref{fig:my_label1} illustrates a typical irradiance profile for a clear-sky day.

\begin{figure*}[ht!]
    \centering
    \includegraphics{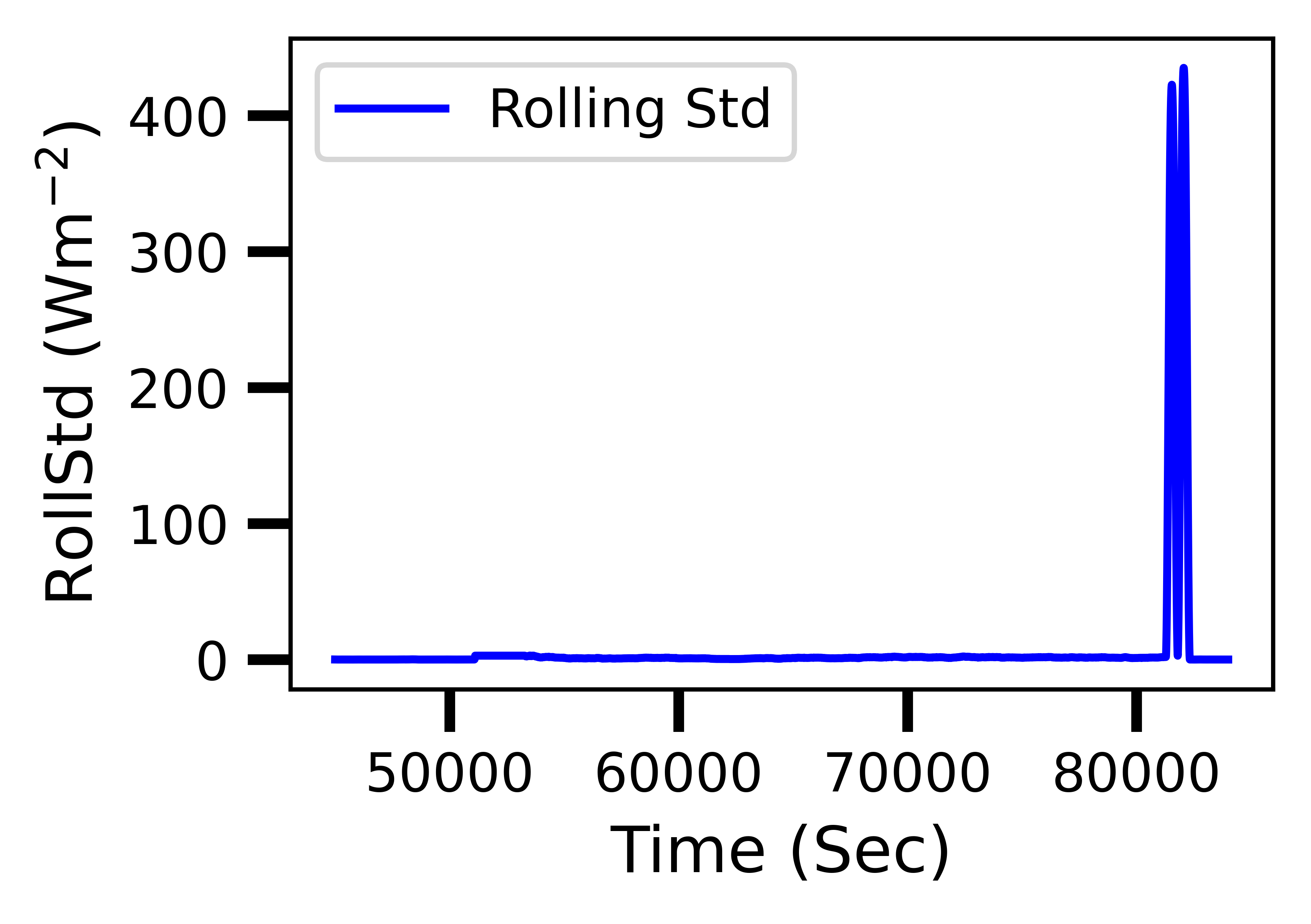}
    \hfill
    \includegraphics{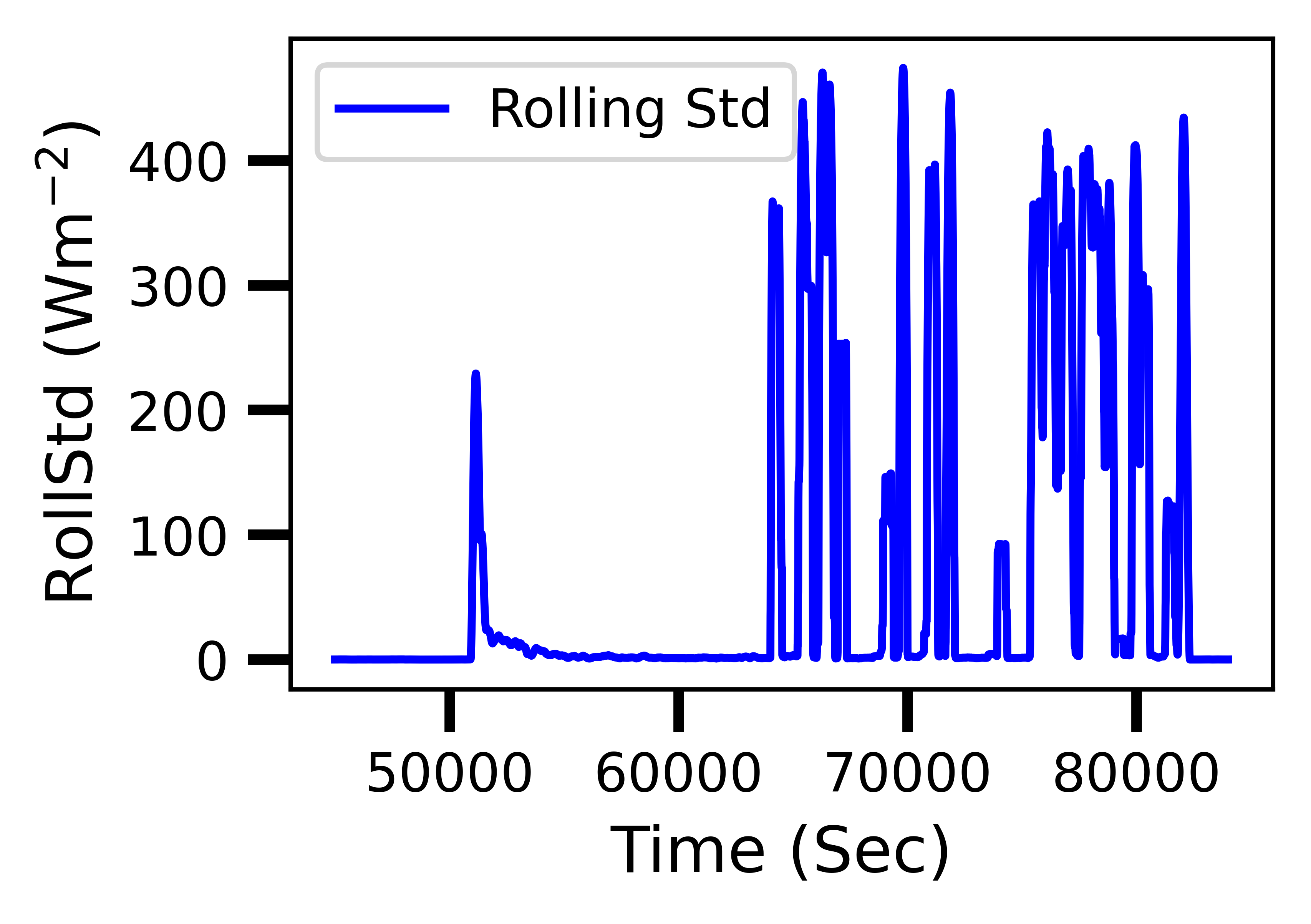} 
    \caption{(a) Figure a shows the rolling standard deviation of the irradiance profile for a clear-sky day, showing consistently low values with a spike at dusk due to the sharp decline in irradiance. 
    (b) Figure b shows the rolling standard deviation of the irradiance profile for a cloudy day, where higher values indicate significant fluctuations in solar irradiance.
    }
    \label{fig:my_lb3}
\end{figure*}

\textit{Clear days} are visually identified for each month across multiple years and interpolated to match the timestamps of a selected \textit{reference day}. The mean of these interpolated clear days forms the monthly \textit{template} (see Figure \ref{fig:my_lb2}). A rolling standard deviation is computed to quantify deviations between each observed day and this template (see Figure \ref{fig:my_lb3}).

To mitigate the influence of solar p-mode oscillations, which have a characteristic period of 5.4 minutes \citep{1988ApJ...324.1158D}, the rolling standard deviation is computed over a 6-minute window (see Section \ref{subsec:3.2}). 

The rolling standard deviation values serve as a quantitative metric for assessing cloud variability. Lower values indicate minimal variation, increasing confidence that the selected observations are free from cloud contamination.

\begin{figure}
    \center
    \includegraphics[width = \columnwidth]{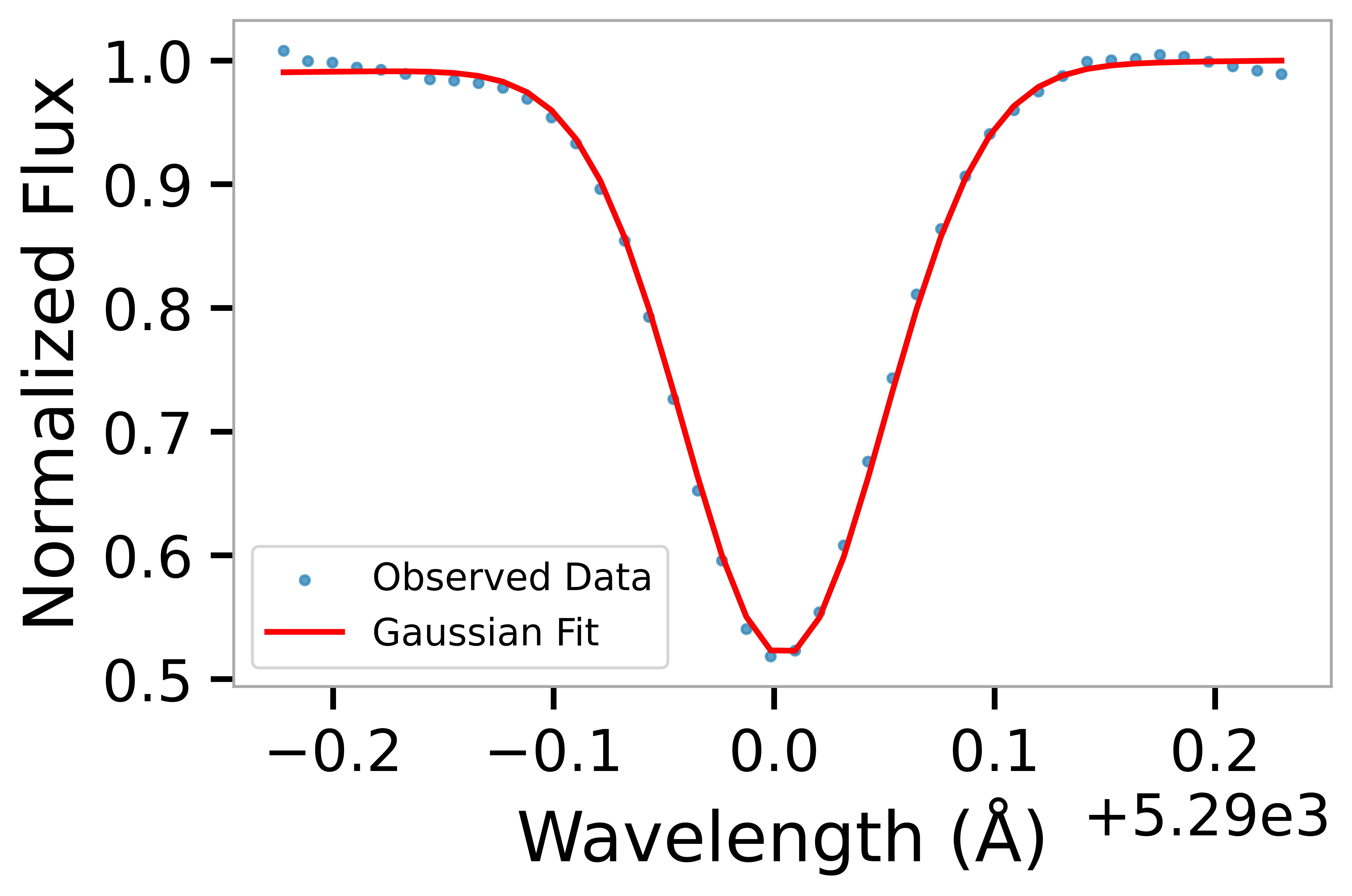}
    \caption{This figure illustrates the Gaussian profile fit to a spectral line, as discussed in Section \ref{subsec:3.3}.}
    \label{fig:my_label4}
\end{figure}

The NEID data files\footnote{\label{neidL2}\href{https://neid.ipac.caltech.edu/docs/NEID-DRP/dataformat.html}{NEID L2 Data Format}} include integration times for each observation. The rolling standard deviation is averaged over the corresponding integration time to assess cloud coverage during these periods. The resulting distribution (see Figure \ref{fig:my_lb2}) has a long tail extending toward higher values. A cutoff of 3 W m$^{-2}$ is chosen to select solar spectral data observed under clear-sky conditions for further analysis.

\subsubsection{Steps to filter remaining outliers}

\label{subsec:3.1.2}

Following clear-sky selection, only High Resolution (HR) mode solar spectra from NEID were retained, and exposures with a signal-to-noise ratio below 300 (as listed in the file headers) were excluded to ensure data quality.

\begin{figure*}
    \center
    \includegraphics[scale = 0.7]{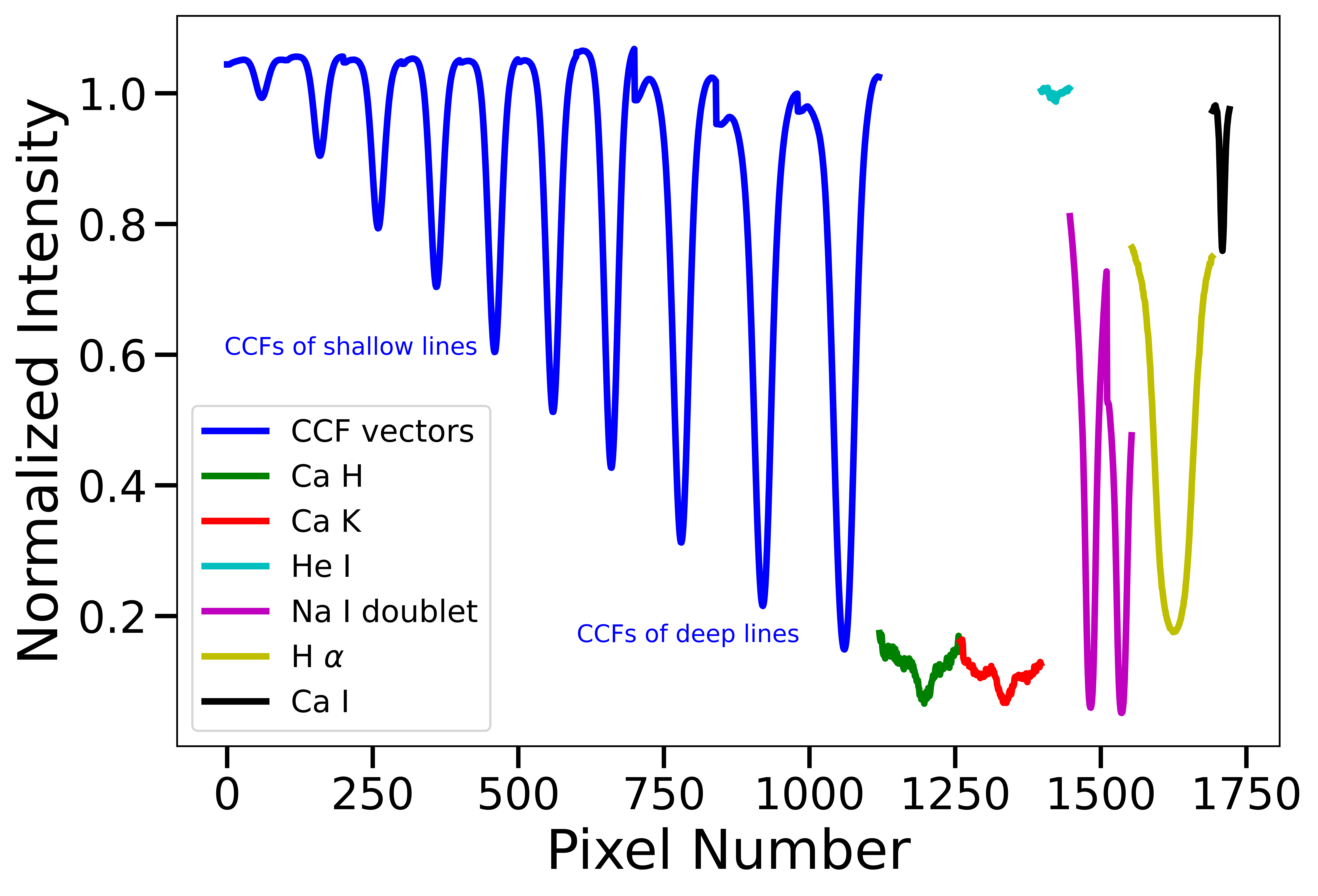}
    \caption{This figure shows a Sample 1D-CCCF vector (see Section \ref{subsec:3.3}) constructed by concatenating 10 CCFs, with activity-sensitive spectral lines as listed in Table \ref{tab:1} stitched together. Activity lines are normalized following the NEID DRP approach \protect\hyperref[neid:Activity]{\textsuperscript{5}}. Each spectrum is represented in this compact form to balance vector size optimization against information loss from averaging.}
    \label{fig:my_label5}
\end{figure*}

Additionally, RV shifts calculated by the NEID DRP\textsuperscript{\ref{neidpipeline}} are analyzed using a histogram. The majority of RVMOD values follow a Gaussian-like distribution. A 3$\sigma$ threshold is applied to remove statistical outliers, eliminating the remaining outlier data points.
  
\subsection{Pre-Processing Steps}

\label{subsec:3.2}

The selected data undergo a series of pre-processing steps to prepare them for machine learning training. These include continuum normalization, heliocentric correction, and temporal averaging to mitigate p-mode oscillations.

First, like all echelle spectrographs, NEID spectra are modulated by the blaze function of the diffraction grating\textsuperscript{\ref{neidpipeline}}. To remove this modulation, the spectra are divided by the effective blaze response across orders, yielding a continuum-normalized spectrum of intensity as a function of wavelength.

Next, a heliocentric correction is applied to account for Doppler shifts caused by Earth's motion and the Sun’s reflex motion due to gravitational interactions with Solar System planets. This correction is computed using the \texttt{Barycorrpy} package \citep{Kanodia_2018}.

Finally, to mitigate the influence of solar p-mode oscillations, we apply temporal averaging. As described in Section~\ref{subsec:3.1.1}, we use a 6-minute rolling window to compute the standard deviation of solar intensity and apply a local average over four consecutive samples to average out p-mode oscillations while preserving the original $\sim$93-second sampling cadence \citep{Lin_2022}.

\begin{figure*}
    \raggedright
    \includegraphics[scale=1.6]{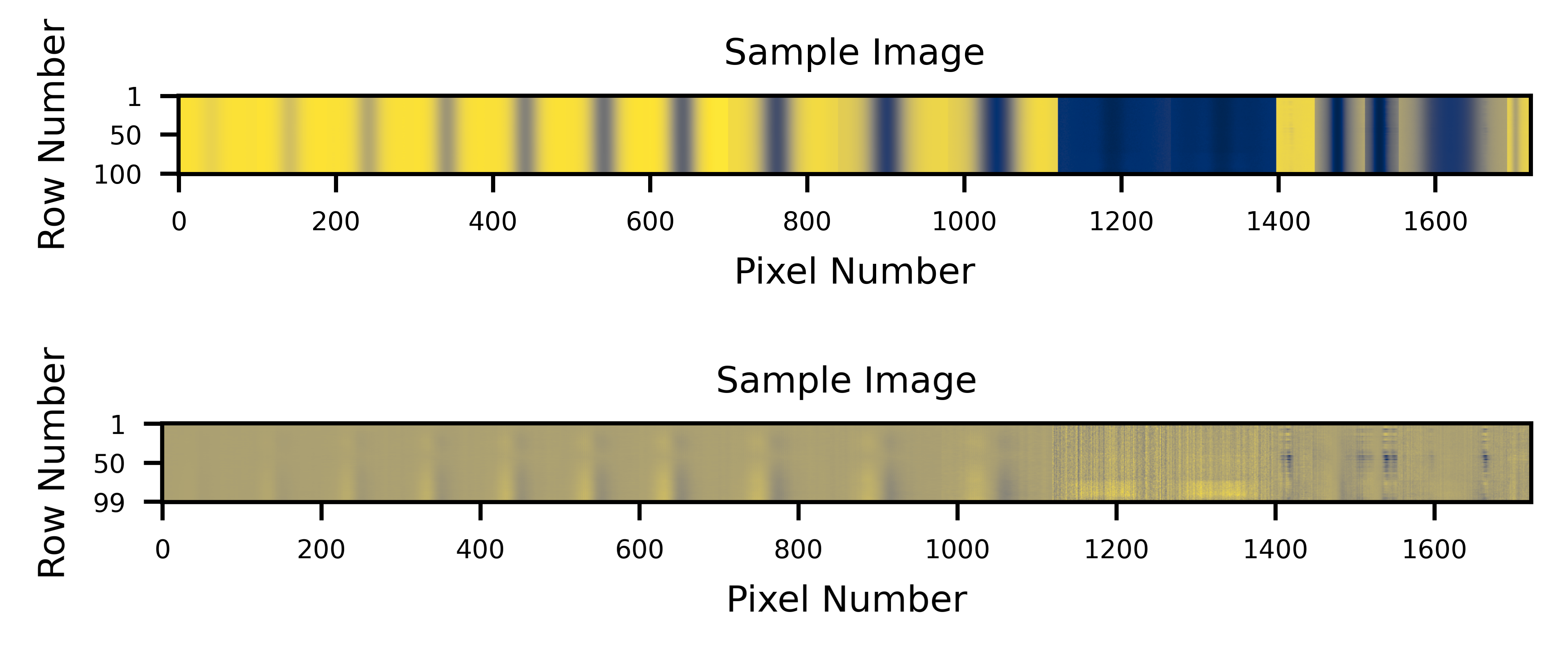}
    \caption{(a) Figure a shows a sample 2D image formed by stacking 100 1D-CCCF vectors, resulting in dimensions of 100 × 1722. (b) Figure b shows a corresponding sample 2D image of CCCF differences, created by subtracting the first row from all subsequent rows, yielding an image of dimensions 99 × 1722. }
    \label{fig:my_label6}
\end{figure*}

\subsection{Generating CCCF vectors} 
\label{subsec:3.3}

The NEID DRP Level 2 dataset\textsuperscript{\ref{neidpipeline},\ref{neidL2}} contains 122 echelle orders, each with 9,216 pixels, resulting in a total of approximately 1.1 million pixels. However, spectral analysis primarily focuses on spectral lines rather than the entire spectrum. The large data volume in these spectral orders presents computational challenges, particularly due to GPU memory limitations in machine learning applications. To mitigate this, the dataset must be reduced in dimensionality while retaining key astrophysical information.

Astrophysically, spectral lines with similar depths originate from comparable heights and temperatures in the photosphere \citep{2020A&A...633A..76C}. To preserve this correlated information, spectral lines are grouped by normalized depth, with minimal blending to ensure accurate profile extraction. Suitable lines are selected using the ESPRESSO G2V line mask\footnote{\href{https://www.eso.org/public/teles-instr/paranal-observatory/vlt/vlt-instr/espresso/}{ESPRESSO database}}, and each line is fitted with a Gaussian profile (see Figure \ref{fig:my_label4}) to evaluate its strength, shape, and blending level.

Additionally, activity-sensitive spectral lines, which are particularly influenced by stellar magnetic or chromospheric activity,  are included (See NEID Documentation\footnote{\label{neid:Activity}\href{https://neid.ipac.caltech.edu/docs/NEID-DRP/algorithms.html\#calculation-of-the-activity-index}{NEID documentation: Stellar Activity info}}). 
Their inclusion improves the model’s ability to isolate planetary signals from stellar activity, enhancing the accuracy of orbital parameter predictions.

To efficiently capture variations in spectral line deformation without losing critical information, the spectral line list is divided into 10 subgroups by partitioning the depth range into evenly spaced bins. Cross-correlation Functions (CCFs) are then computed separately for each subgroup, preserving subtle differences in spectral lines that would be averaged out in a global CCF.

The resulting CCFs exhibit a well-defined flat continuum with a central dip, representing averaged spectral line profiles. The velocity axis spans from -200 to 201 kms$^{-1}$, sampled at 1604 pixels. To reduce noise and focus on the relevant velocity range, the CCFs are symmetrically trimmed around the central dip. A 100-pixel window (-12.5 to 12.5 kms$^{-1}$) is applied to the first seven CCFs, while the remaining three, corresponding to broader and deeper spectral lines, are trimmed using a 140-pixel window (-17.5 to 17.5 kms$^{-1}$). The final set of 10 trimmed CCFs is concatenated into a single vector of length 1120.

Finally, the activity-sensitive spectral lines are appended to these concatenated vectors, resulting in Concatenated Cross-Correlation Function (1D-CCCF) vectors, each with a total length of 1722 pixels (see Figure \ref{fig:my_label5}). Table \ref{tab:1} details the properties of the activity-sensitive spectral lines.

These 1D-CCCF vectors serve as the foundational input units for generating synthetic time-series datasets, which are structured into 2D-CCCF representations used in our model training pipeline, as detailed in the following section.

\begin{figure*}
    \centering
    \includegraphics[scale=0.5]{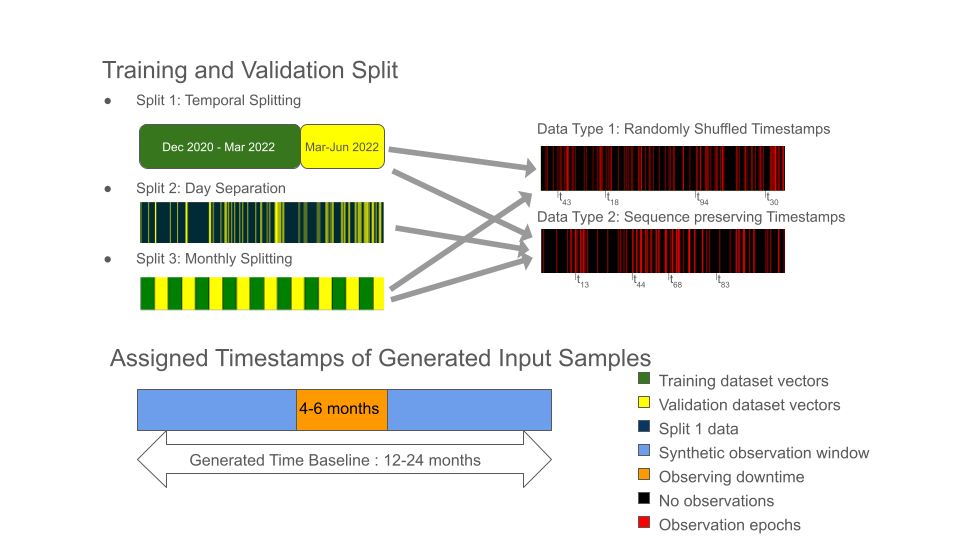}
    \caption{
    This figure illustrates how observational data are partitioned into training and validation sets, and how individual samples are formatted for model input.
    (a) The left side illustrates how the dataset is partitioned across multiple validation strategies. In Split 1, validation subsets V1 and V2 are selected using a time-contiguous strategy. Split 2 is applied concurrently, where the validation subset V3 is defined by day-separated observations. In this Split 2 view, the remainder of the dataset, corresponding to the Split 1 training and validation regions, is dimmed (dark blue) to highlight the distinct structure of V3. \\    A monthly-based split is also used, enabling training over longer timescales while preserving a distinct validation set M (see Section~\ref{subsec:4.2}). The right side depicts the format of individual data samples. Each sample is a sequence of 100 epochs, where colored regions indicate epochs with observations and black regions mark epochs with no observation. Two formats are used; one that preserves the temporal order of observed epochs, and another where observations are randomly shuffled to remove sequential information.\\
    (b) Observation uptime and downtime: observation baselines vary between 12–24 months, interspersed with typical downtime periods lasting 4–6 months due to mission scheduling or survey gaps.
    }
    \label{fig:my_label7}
\end{figure*}

\begin{table}[h]
    \centering
    \caption{This table lists the activity-sensitive spectral Lines used in CCCF vector generation}
    \label{tab:1}
    \begin{tabular}{ccc}
        \toprule
        \textbf{Index} & \textbf{Line Center(\r{A})} & \textbf{Line Width(\r{A})}  \\
        \midrule
        Ca II H & 3968.470 & 1.09 \\
        Ca II K & 3933.664 & 1.09 \\
        He I & 5875.62 & 0.4 \\
        Na I & 5895.92 & 0.5 \\
        Na I & 5889.95 & 0.5 \\
        H $\alpha$ & 6562.808 & 0.6 \\
        Ca I & 6572.795 & 0.34\\
        \bottomrule
    \end{tabular}
\end{table}

 \section{Data generation procedure}
 \label{sec:new4}

Extracting orbital parameters from radial velocity spectra requires training data that reflect both astrophysical signals and the irregularities of real-world observations. To this end, we construct time-series datasets from the 1D-CCCF vectors introduced in Section~\ref{subsec:3.3}, with each sample designed to approximate the duration and sampling variability of a realistic observing sequence. This section outlines our approach to injecting Keplerian signals and assembling datasets for model training.

A time series of 100 epochs is constructed by randomly selecting 100 1D-CCCF vectors, each containing 1722 pixels, from the observation period. Two formats are used: in one, the 1D-CCCF vectors are retained in their original temporal order, preserving the irregular cadence of real observation timestamps, while in the other, the 1D-CCCF vectors are randomly shuffled and assigned synthetic timestamps spanning 1 to 2 years. In both cases, the resulting data is organized into a 2D matrix (2D-CCCF vector) where each row corresponds to a single 1D-CCCF vector (see Figure~\ref{fig:my_label6}). To mimic real observing conditions, each of these 2D-CCCF vectors includes an observational downtime of 4 to 6 months per year, implemented by masking out a continuous block of dates randomly centered across the year (see Figure~\ref{fig:my_label7}).

A Keplerian signal for an elliptical orbit is sequentially injected into each row based on its assigned timestamp, using the \texttt{radvel} \citep{Fulton_2018} toolkit. 
The orbital parameters for the injected Keplerian orbit were selected from the following ranges:

\begin{itemize}
    \item Orbital period($P$): 12–365 days
    \item Semi-amplitude($K$): 0.05–3 ms$^{-1}$
    \item Eccentricity($e$): 0–0.6 
    \item Argument of periastron($\omega$): 0–2$\pi$ 
\end{itemize}

The period $P$ is sampled uniformly in logarithmic space within its specified range, while the remaining orbital parameters follow uniform distributions across their respective ranges. Consequently, the resulting 2D-CCCF vector consists of rows with varying Doppler shifts, each corresponding to a different time for the same Keplerian signal.

\begin{figure*}
    \center
    \includegraphics[scale=0.7]{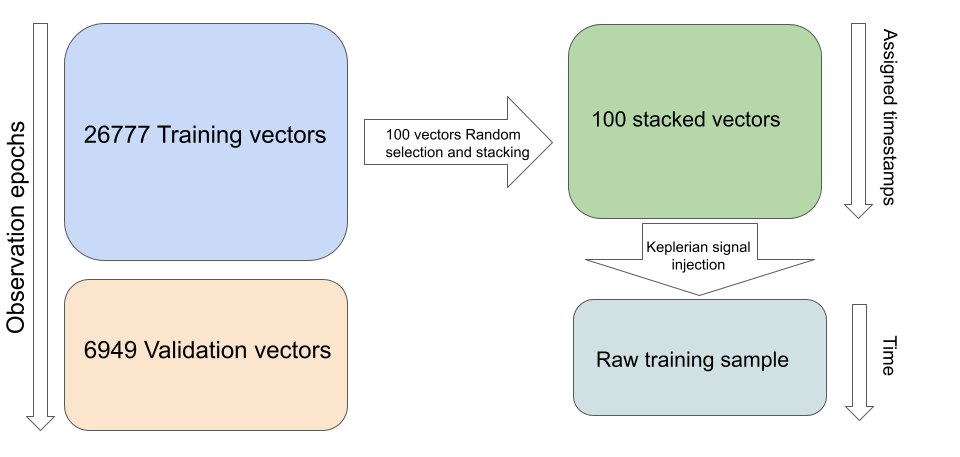}
    \caption{This figure provides a schematic representation of the generation of temporally shuffled data samples. The left section shows the disjoint training and validation time spaces, while the bottom-right section depicts the generated raw training sample. Day-separated validation data (set V3, see Figure \ref{fig:my_label7}) are excluded from the training dataset.}

    \label{fig:my_label8}
\end{figure*}

To emphasize variations between observations, the difference between each row of the 2D-CCCF vector and the first row vector is computed. This transformation reduces the number of rows to 99 while preserving the essential dynamical information. The processed dataset is then used as input for the training algorithm. A schematic representation of this procedure is shown in Figure \ref{fig:my_label8}.

The CCCF vectors are influenced by two primary effects: the applied Doppler shift, which induces periodic spatial translation, and intrinsic stellar activity, which causes both translational shifts and structural distortions in the CCCF profile \citep{2020A&A...633A..76C}. To accurately recover the periodic Doppler signal, the model must distinguish between these two effects.

\begin{figure}
    \centering
    \includegraphics[width=\columnwidth]{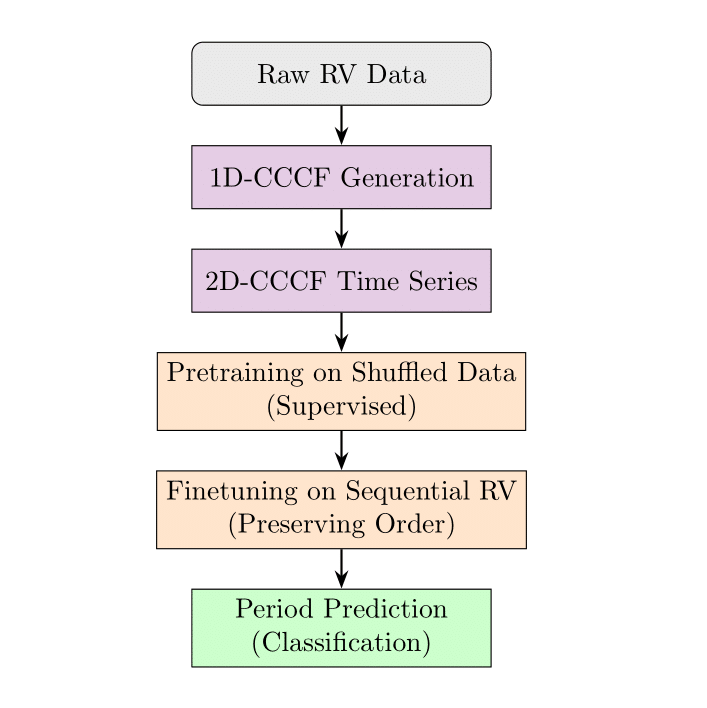}
    \caption{
    This figure illustrates a schematic workflow of our Machine Learning pipeline for RV-based period prediction. The process begins with raw RV data, which is transformed into 1D concatenated cross-correlation functions (1D-CCCFs) and further stacked to form 2D concatenated cross-correlation functions (2D-CCCFs) that serve as input representations. Supervised pretraining is performed on temporally shuffled data to enable the model to learn generic Keplerian Doppler shift signatures independent of temporal correlations. This is followed by fine-tuning on sequential RV observations to expose the model to realistic temporal stellar activity patterns. The trained model then performs classification-based coarse prediction of the Keplerian period corresponding to the sought planetary signal.
    }
    \label{fig:flowchart}
\end{figure}

The processing steps described above produce a 2D image comprising 100 observations of a single Sun-planet system. The machine learning model is trained to extract the system’s orbital parameters from these spectral representations.

\begin{figure*}
    \center
    \includegraphics[scale=0.4]{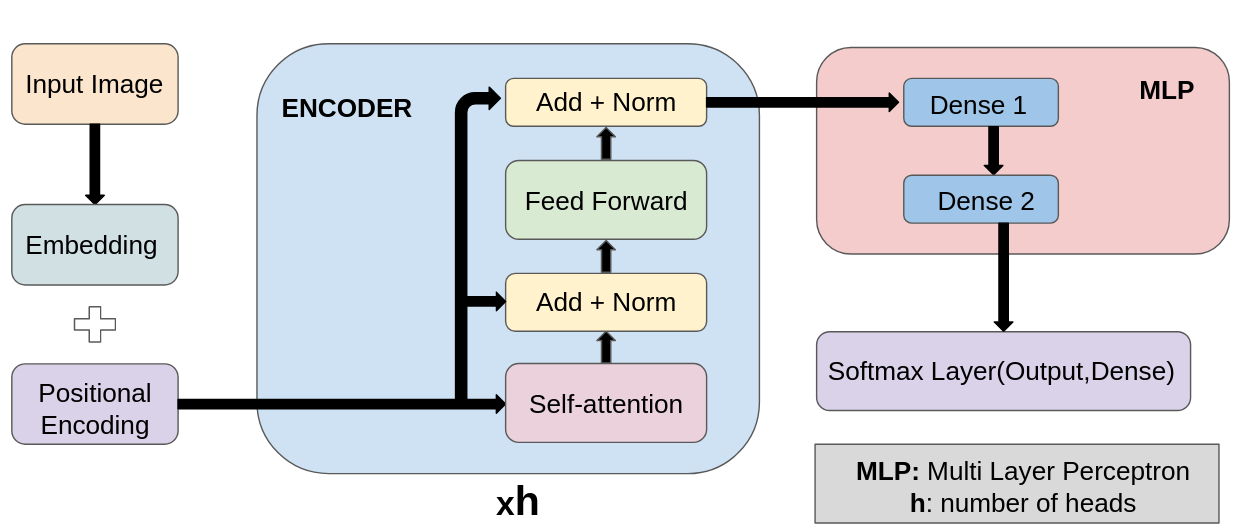}
    \caption{This figure presents the architecture of our Vision Transformer (ViT) model. Input images are divided into patches, embedded, and subsequently augmented with positional encodings. The encoder processes these embeddings using self-attention mechanisms and a multilayer perceptron (MLP). The final output is a parameter likelihood vector generated via a softmax layer.}
    \label{fig:my_label9}
\end{figure*}


\subsection{Dataset Splitting}
\label{subsec:4.1}

After pruning, processing, and constructing the 1D-CCCF dataset as described in Section~\ref{sec:3}, we obtain a total of 35,757 vectors. To ensure robust model evaluation across both shuffled and temporally ordered conditions, this dataset is first split into distinct training and validation subsets. Each subset is then independently processed into 2D-CCCF representations using the Keplerian injection method described in Section~\ref{sec:new4}, ensuring consistent generation across all partitions. The full list of the 35,757 filenames used in this dataset is available on Github\footnote{\label{ClearSkyNEIDSolarSpectraList}\href{https://github.com/ADG97astro/ClearSkyNEIDSolarSpectraList}{ClearSkyNEIDSolarSpectraList}}. The splitting strategies are illustrated in Figure~\ref{fig:my_label7}:

\begin{itemize}

\item Split 1 (top panel) defines the primary training–validation separation, where 26,777 1D-CCCFs are used for training and 6,949 1D-CCCFs for validation. This split is temporally disjoint, ensuring that data from the same observing nights are not shared between the two subsets.

\item Split 2 (bottom panel) corresponds to a separately reserved validation subset comprising 2,031 1D-CCCF vectors drawn from entirely distinct observing days not present in either the training or validation sets from Split 1. This \enquote{day-separated} validation data spans the full dataset duration, enabling evaluation of the model's ability to generalize across varying time baselines.

\end{itemize}

Each of these subsets is independently processed into 2D-CCCF representations (see Section \ref{sec:new4}). Thus, Splits 1 and 2 represent two facets of the same overall partitioning strategy: Split 1 supports standard training and validation, while Split 2 enables robust cross-epoch evaluation on non-overlapping days.

\subsection{Training and Validation Data}
\label{subsec:4.2}

From the processed training and validation vector sets (see Section \ref{subsec:4.1}), we generate 840,000 training samples and 500,000 validation samples (see Section \ref{sec:new4}). The validation samples are categorized into three distinct sets based on their sampling methodology and timestamp assignment:

\begin{itemize}
    \item \textsc{Set V1}: Validation samples with randomly assigned timestamps, derived from Split~1. These are generated using the same methodology as the training dataset (see Figure \ref{fig:my_label8}). 
    \item \textsc{Set V2:} Validation samples derived from Split~1, preserving the chronological order of raw spectral observations, but with timestamps rescaled to match the training data distribution.  
    \item \textsc{Set V3:} Validation samples with unmodified timestamps, derived from a separate dataset (Split 2) spanning the full 19-month observation period (see Section \ref{subsec:4.1}, Figure~\ref{fig:my_label7}).  
\end{itemize}

To distribute the planned 2000–2500 validation samples in set V3 more broadly across the timeline, we prioritized days with fewer retained observation epochs. This allowed the limited validation set (2031 samples) to span a wider range of epochs while preserving sufficient training data. The varying gaps between validation segments (Figure~\ref{fig:my_label7}, Split 2) reflect the fact that days with fewer observations, after the filtering procedures described in Section~\ref{sec:3}, are unevenly distributed in time.

By employing multiple validation sets, we ensure a robust assessment of model performance across different sampling strategies and temporal distributions.

In addition to the primary dataset partitioning strategy, an alternative Split 3 is implemented to train an additional model (see Figure \ref{fig:my_label7}). Rather than a single temporal division, the training and validation datasets are segmented by calendar months: observations from odd-numbered months are assigned to the training set, while those from even-numbered months are allocated to the validation set \enquote{M}. To further reduce temporal correlations, data from the first and last two days of each month are excluded. This partitioning strategy extends the temporal coverage of the validation set, increasing variability within the ordered samples compared to the previous approach, where set V2 is scaled. Despite these differences, all subsequent processing steps remain identical across both shuffled and ordered datasets, and no additional scaling is applied. The final training and validation datasets span approximately 18 months, with systematic monthly gaps throughout the year.

\section{Training Procedure}
\label{sec:4}

With a diverse and carefully partitioned dataset in place, we now describe the model architecture and training strategy used to extract the underlying Keplerian parameters.

We train deep learning models to infer orbital parameters, specifically the period and semi-amplitude, from time-series representations of spectral 1D-CCCF vectors (see Section \ref{sec:new4}). These 2D-CCCF vectors are normalized, and the outputs are transformed into parameter likelihood vectors for each orbital parameter. The model learns to map these inputs to their respective outputs, with performance evaluated across multiple datasets to assess generalizability. While the overall network architecture remains unchanged for both parameters, the output dimensionality varies based on the length of the parameter likelihood vectors.

To systematically understand and test the model’s ability to extract Keplerian signals from observational data, we adopt a two-stage training strategy. In the first stage, the model is trained on datasets with randomized observation timestamps, which removes temporal coherence while preserving the overall scatter in the radial velocities. This shuffling is not merely a data augmentation step but a design choice that ensures the model cannot overfit time-correlated variability. Instead, it must learn to recognize the underlying Doppler transformation due to orbital motion within a noisy background, separate from temporally correlated stellar activity. 

In the second stage, the model is fine-tuned on an ordered dataset with realistic time sampling, which reintroduces temporal coherence reflective of actual observational conditions.

Training directly on temporally ordered data was found to cause the model to overfit to sampling artifacts or activity-driven variability, reducing its ability to generalize. By contrast, the two-stage setup, starting from shuffled inputs, forces the model to first learn the underlying Keplerian Doppler shifts. The fine-tuning stage then allows the model to adjust to realistic conditions without overriding the core Keplerian signal representations.

This stepwise introduction enables the model to learn how time-dependent activity patterns influence signal recovery, bridging the gap between randomized and real-world sampling. To illustrate the model’s practical applicability, we apply it to a Sun-planet system using 100 aperiodically sampled spectral observations.

Figure \ref{fig:flowchart} shows the overall workflow of our ML pipeline, from initial RV data preprocessing through pretraining and finetuning, to the final stage of coarse period classification (see Figure~\ref{fig:sample_output}).

For consistency, periodogram comparisons are performed on both randomized and ordered versions of the dataset, ensuring a fair evaluation by using the same data instances for both the traditional periodogram and the machine learning model in each configuration.

\subsection{Model Architecture}
\label{subsec:4.3}

We use Vision Transformers (ViTs), a variant of the Transformer model \citep{vaswani2023attention}, to analyze RV time-series data. The Transformer architecture employs self-attention mechanisms to assign varying importance to different input components, enabling it to capture both short- and long-range dependencies, which is essential for accurately extracting RV signals.

Originally developed for image processing, ViTs represent input images as sequences of patches. In our approach, each row of a 2D-CCCF vector, corresponding to a shifted 1D-CCCF vector, is treated as a patch, allowing the model to capture both spectral and temporal information effectively. These patches are flattened and transformed into contextual embeddings, which encode relevant features in a reduced-dimensional space, improving model accuracy while reducing computational complexity. Positional encodings are incorporated into the patch embeddings to retain spatial and temporal relationships, which the original Transformer architecture does not inherently capture due to its non-sequential nature.

The continuous parameter space of orbital period and semi-amplitude is discretized for classification. The orbital period is divided logarithmically into 10 bins labeled 0 to 9, while the semi-amplitude is segmented into 5 equal linear bins labeled 0 to 4. Preliminary experiments using a regression formulation were found to be unstable, particularly at low SNR. We therefore adopt a classification approach, which consistently led to better convergence and accuracy (see Appendix \ref{appendix:classvsreg} for details).

This reformulation improves model stability, provides a measure of uncertainty through the predicted probability distribution, and enhances the model’s ability to distinguish between different parameter ranges.

The ViT architecture employs multiple self-attention heads to capture diverse attention patterns, enabling the extraction of complex spectral and temporal relationships. The outputs from these attention heads are concatenated, linearly transformed, and mapped to discrete probability distributions over the orbital parameters.

Additionally, the architecture supports generalization and transfer learning, allowing fine-tuning on datasets from other stars, provided the model is pre-trained on a sufficiently diverse set of solar RV observations. A detailed schematic of our architecture is presented in Figure \ref{fig:my_label9}.

To ensure effective optimization, the cross-entropy loss function, well-suited for multi-class classification tasks, is employed to measure discrepancies between predicted and true distributions. Stochastic Gradient Descent (SGD) is used as the optimizer, with the learning rate set to $10^{-3}$. The loss contributions from orbital period and semi-amplitude predictions are weighted equally to maintain balanced optimization across both parameters.

The model is trained to differentiate between activity-induced variations and Keplerian RV shifts, thereby improving its predictive accuracy for orbital parameters. The final model, selected based on the lowest validation loss, is retained for future applications.

\begin{figure}
    \center
    \includegraphics[width = \columnwidth]{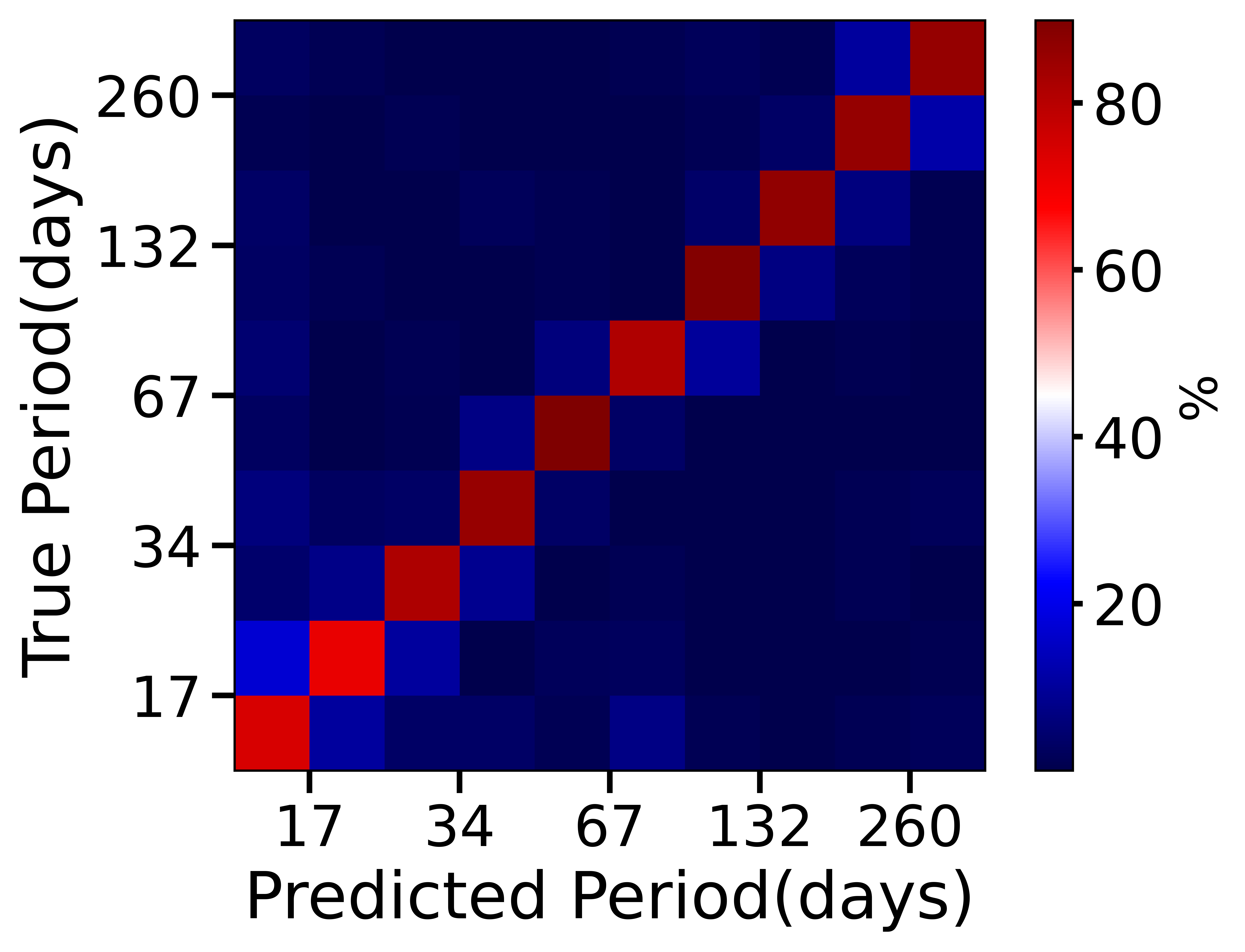}
    \caption{This confusion matrix illustrates the model's performance in predicting orbital periods on the shuffled dataset V1. The matrix is normalized along the \enquote{True Period} axis for each bin. Predictions are concentrated along the diagonal, reflecting high overall accuracy, with slightly reduced accuracy observed in the lower period bins.}
    \label{fig:my_lb10}
\end{figure}

\section{Results}
\label{sec:5}

\subsection{For temporally shuffled data}
\label{subsec:5.1}

In this study, the training dataset was constructed in a manner similar to the validation set V1 (see Figure \ref{fig:my_label8}), although the datasets are temporally distinct, implying minimal inherent correlation between them. As a result, the model achieves prediction accuracies of 86\% for the orbital period and 76\% for the semi-amplitude when tested on the V1 validation set.

\subsubsection{Orbital Period}
\label{subsec:5.1.1}

\begin{figure}
    \center
    \includegraphics[width = \columnwidth]{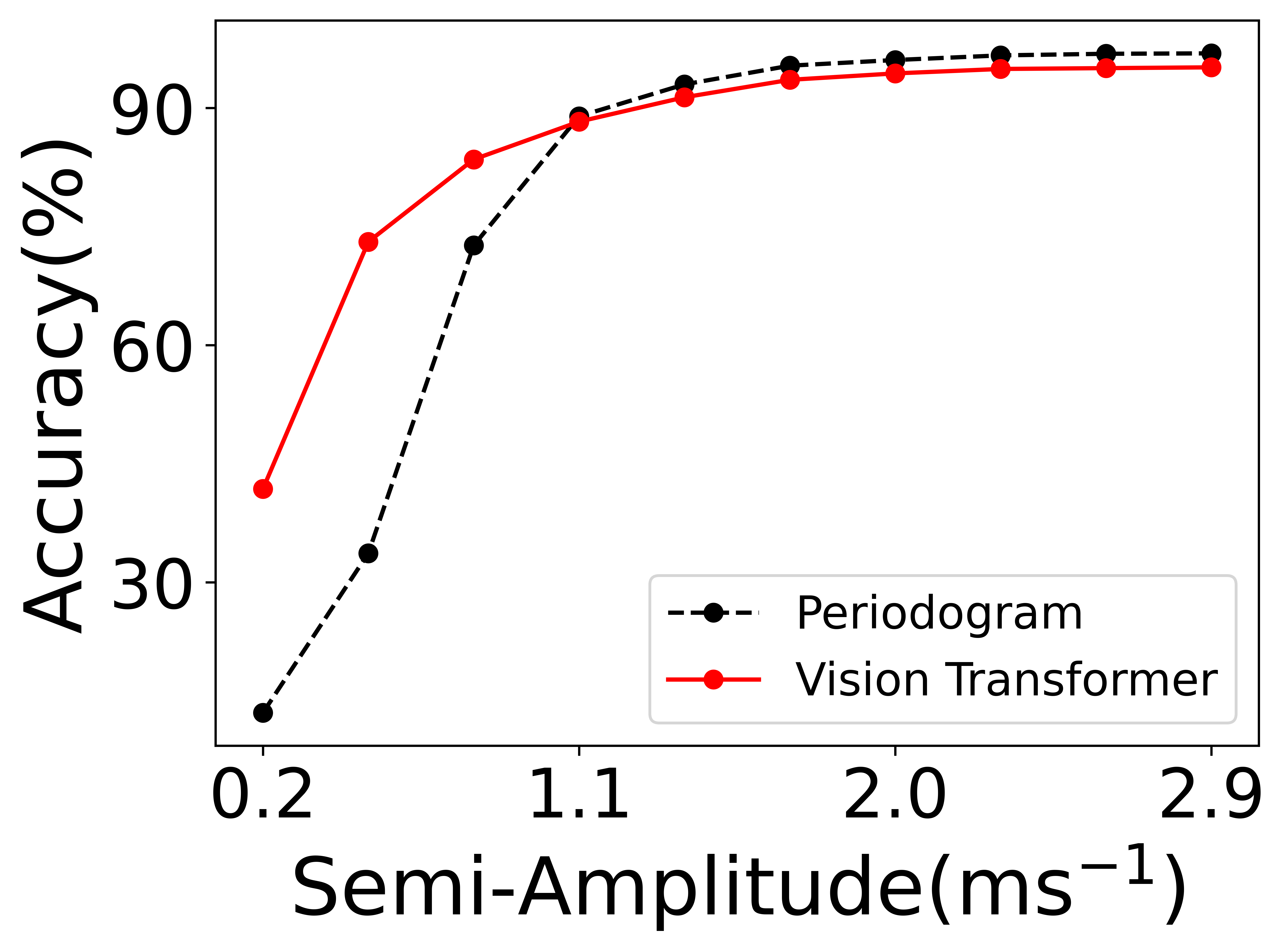}
    \caption{This figure shows a comparison of the Lomb-Scargle periodogram and machine learning model performance for the shuffled dataset V1. The figure compares how accurately each method identifies the correct period bin, with the periodogram’s power spectrum maxima discretized to align with the bin structure of the machine learning model, enabling a direct comparison. The model achieves significantly higher accuracy at lower semi-amplitudes, while the periodogram slightly surpasses the model by approximately 3\% at higher amplitudes. This comparison focuses on discretized outputs, excluding factors such as peak amplitude and false alarm probabilities inherent to the periodogram.}
    \label{fig:my_lb11}
\end{figure}

\begin{figure}
    \center
    \includegraphics[width = \columnwidth]{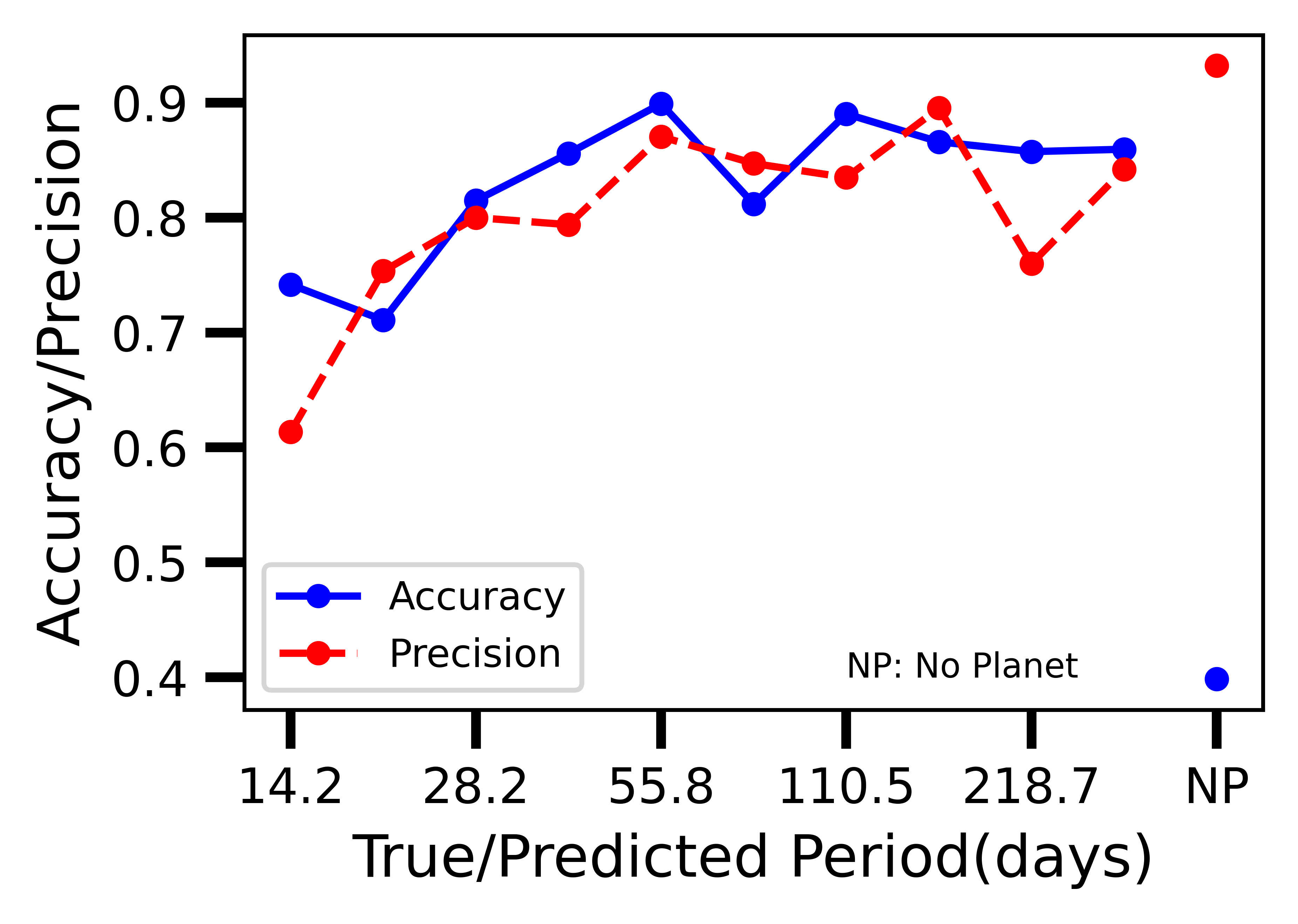}
    \caption{This figure illustrates the model's accuracy and precision for the shuffled dataset V1, highlighting its performance across different period labels. Accuracy indicates the fraction of correctly predicted cases for each true period label, while precision represents the proportion of predictions for a given label that correspond to true positives. Both metrics generally show strong alignment across most period values, with two notable exceptions: the shortest period class and the \enquote{no planet} hypothesis (last label). In the \enquote{no planet} scenario, high precision demonstrates that the model’s no-planet predictions are largely correct. However, low accuracy reveals frequent misclassification of true no-planet cases as planetary detections. In contrast, the shortest period class exhibits a less pronounced but opposite effect, where accuracy surpasses precision, leading to a divergence between the two metrics in these specific scenarios.}
    \label{fig:my_label12}
\end{figure}

\begin{figure}
    \center
    \includegraphics[width = \columnwidth]{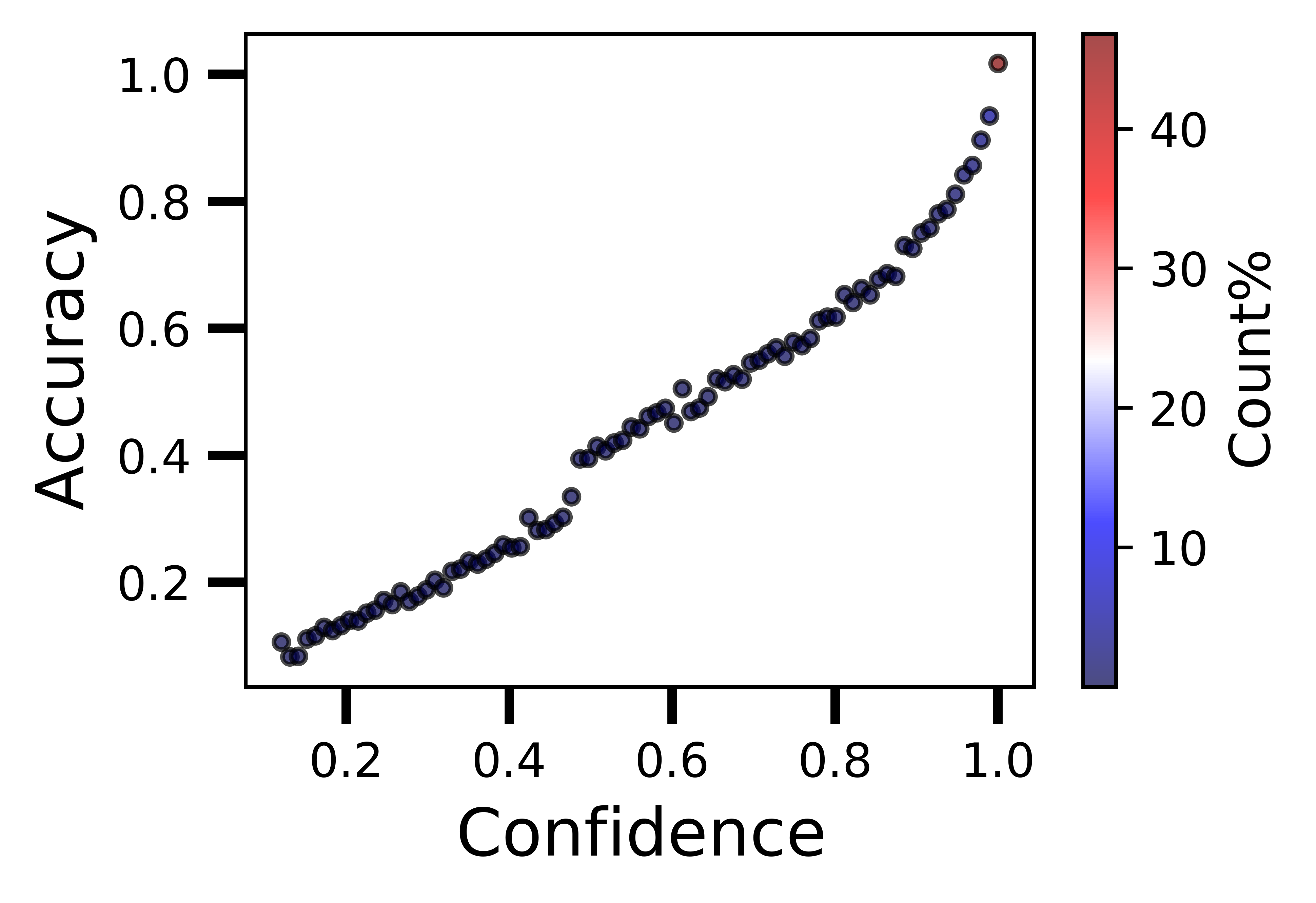}
    \caption{This plot depicts the relationship between period prediction confidence and accuracy for the shuffled dataset V1. The confidence values, ranging from 0 to 1, are divided into 100 bins, with the corresponding accuracy values depicted for each bin. As confidence in the machine learning model's period predictions increases, accuracy improves. Low-confidence predictions exhibit minimal accuracy, while accuracy steadily increases and approaches 1 as confidence nears its maximum value. Notably, approximately 40\% of the predictions fall into the highest confidence bin (confidence $>$ 0.99), where accuracy reaches nearly 99\%. This distribution suggests a saturation effect, with a significant accumulation of predictions in the highest confidence range.}

    \label{fig:my_lb13}
\end{figure}

\begin{figure}
    \centering
    \includegraphics[width = \columnwidth]{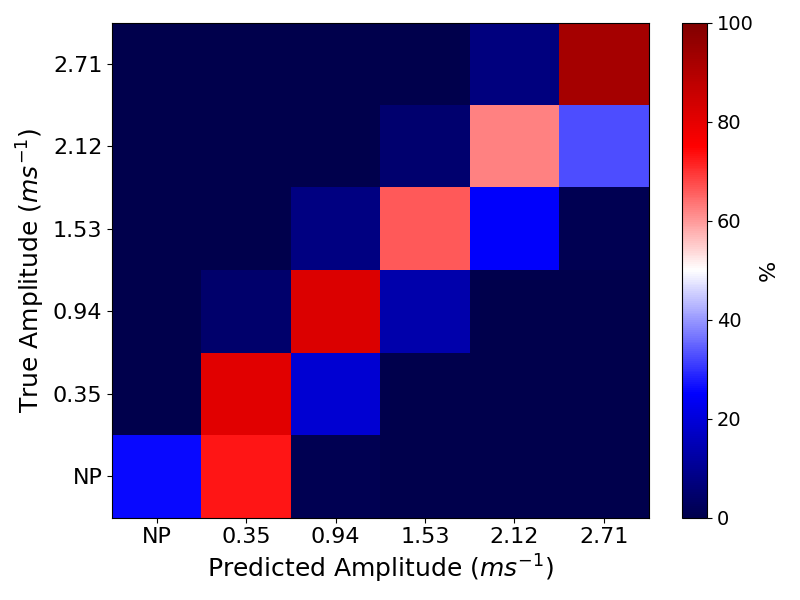}
    \caption{This confusion matrix illustrates the model's performance in predicting semi-amplitudes on the shuffled dataset V1. The matrix is normalized along the \enquote{True Semi-Amplitude} axis for each bin, and includes the \enquote{No Planet} (NP) scenario.  }
    \label{fig:my_lb14}
\end{figure}

\begin{figure}
    \centering
    \includegraphics[width = \columnwidth]{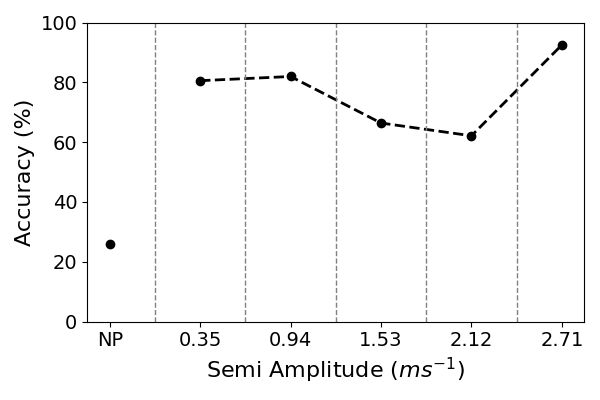}
    \caption {This figure shows the classification accuracy for semi-amplitude predictions in the shuffled dataset V1. The model uses a six-class scheme: five linearly spaced bins representing increasing planetary semi-amplitudes, along with a separate \enquote{No Planet} (NP) category. The model achieves an overall accuracy of 76\% for all planetary systems. Accuracy is highest for the first two lowest amplitude bins and the highest amplitude bin, and decreases across the intermediate bins. The NP scenario shows notably lower accuracy, with many instances misclassified into the lowest amplitude bin.
    }  

    \label{fig:my_lb15}
\end{figure}

Our model accurately predicts orbital period bins in the temporally shuffled dataset V1, demonstrating strong performance in both training and validation. When the activity-sensitive spectral lines previously appended to the CCFs (see Section \ref{subsec:3.3}) are excluded from the CCCF representation, a modest drop in overall accuracy (about 5\%) is observed, indicating that these features provide useful contextual information for identifying planetary periodicities.

Figure \ref{fig:my_lb10} presents the confusion matrix for these period predictions. The model’s high accuracy is evident from the concentration of correctly classified values along the diagonal of the confusion matrix.

The injected Keplerian signals are grouped into 10 linearly spaced semi-amplitude intervals to analyze performance trends. Within these intervals, period bin prediction accuracy increases systematically with semi-amplitude (see Figure \ref{fig:my_lb11}).

The accuracy in the first bin, corresponding to the lowest semi-amplitude values, is approximately 40\% and increases to $\approx$ 94\% for the highest bins. This performance exceeds that of periodogram-based predictions for the same observations at semi-amplitudes up to 1 ms$^{-1}$ (see Figure \ref{fig:my_lb11}, Section \ref{subsec:5.5}).

Figures \ref{fig:my_label12} and \ref{fig:my_lb13} illustrate the variation in period prediction accuracy with orbital period and confidence scores, respectively. Figure \ref{fig:my_label12} also includes the \enquote{No Planet} scenario, where accuracy and precision exhibit distinct behavior (see figure captions for details). The trend of increasing accuracy with confidence, seen in Figure \ref{fig:my_lb13}, indicates that predictions made with higher confidence (defined as the model’s assigned probability to the predicted period bin) are statistically more accurate.

\subsubsection{Semi-amplitude}
\label{subsec:5.1.2}

Our semi-amplitude predictions exhibit strong performance, achieving an overall accuracy of 76\% using a five-bin linear classification scheme for planetary systems. The accuracy trend reveals a distinct trend (see Figures \ref{fig:my_lb14}, \ref{fig:my_lb15}), where the lowest two and highest amplitude bins are predicted with greater accuracy than intermediate bins. The \enquote{No Planet} scenario is predicted with much poorer accuracy, with most misclassifications predicting the lowest amplitude bin.

\subsection{For Temporally Ordered Data}
\label{subsec:5.2}

Our machine learning model effectively predicts orbital parameters in validation set V1. However, in validation sets V2 and V3, where temporal order is preserved (see Section \ref{subsec:4.2}), the model frequently misidentifies solar rotation as the dominant periodic signal, leading to incorrect predictions of the true Keplerian signal.

\begin{figure}
    \center
    \includegraphics[width = \columnwidth]{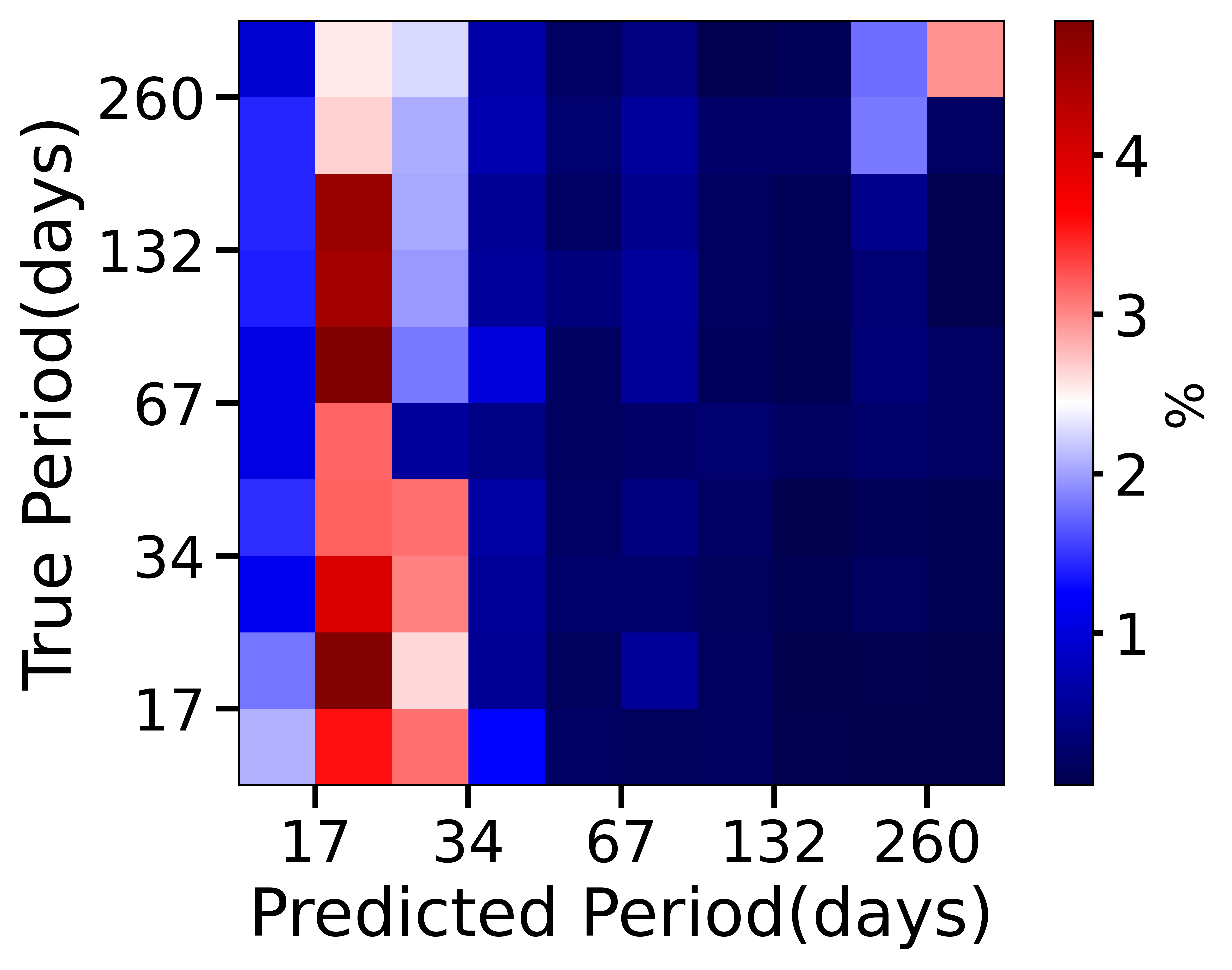}
    \caption{This confusion matrix illustrates the model's performance in predicting orbital periods for the ordered dataset V3 without fine-tuning. The vertical band near 25 days reflects the model's strong bias toward predicting the solar rotation rate, underscoring the need for fine-tuning. Unlike other matrices in this study, this matrix is not normalized along the \enquote{True Period} axis.}  

    \label{fig:my_lb16}
\end{figure}

The impact of this issue differs between V2 and V3. In the scaled sample set V2, the effect is mitigated, likely because most samples correspond to systems with longer orbital periods, resulting in fewer observed stellar rotation cycles per sample. Conversely, the unscaled and more realistic samples in V3 exhibit stronger contamination, with model predictions clustering near the solar rotation period of approximately 25 days (see Figure \ref{fig:my_lb16}). 

Given the limited number of unique ordered data sequences available in the NEID dataset we have utilized, directly training the model on this subset risks overfitting. To address this, we adopt a fine-tuning approach that enhances the model’s ability to differentiate between Keplerian signals and stellar rotation, the two dominant periodic components in the data.

\begin{figure}
    \center
    \includegraphics[width = \columnwidth]{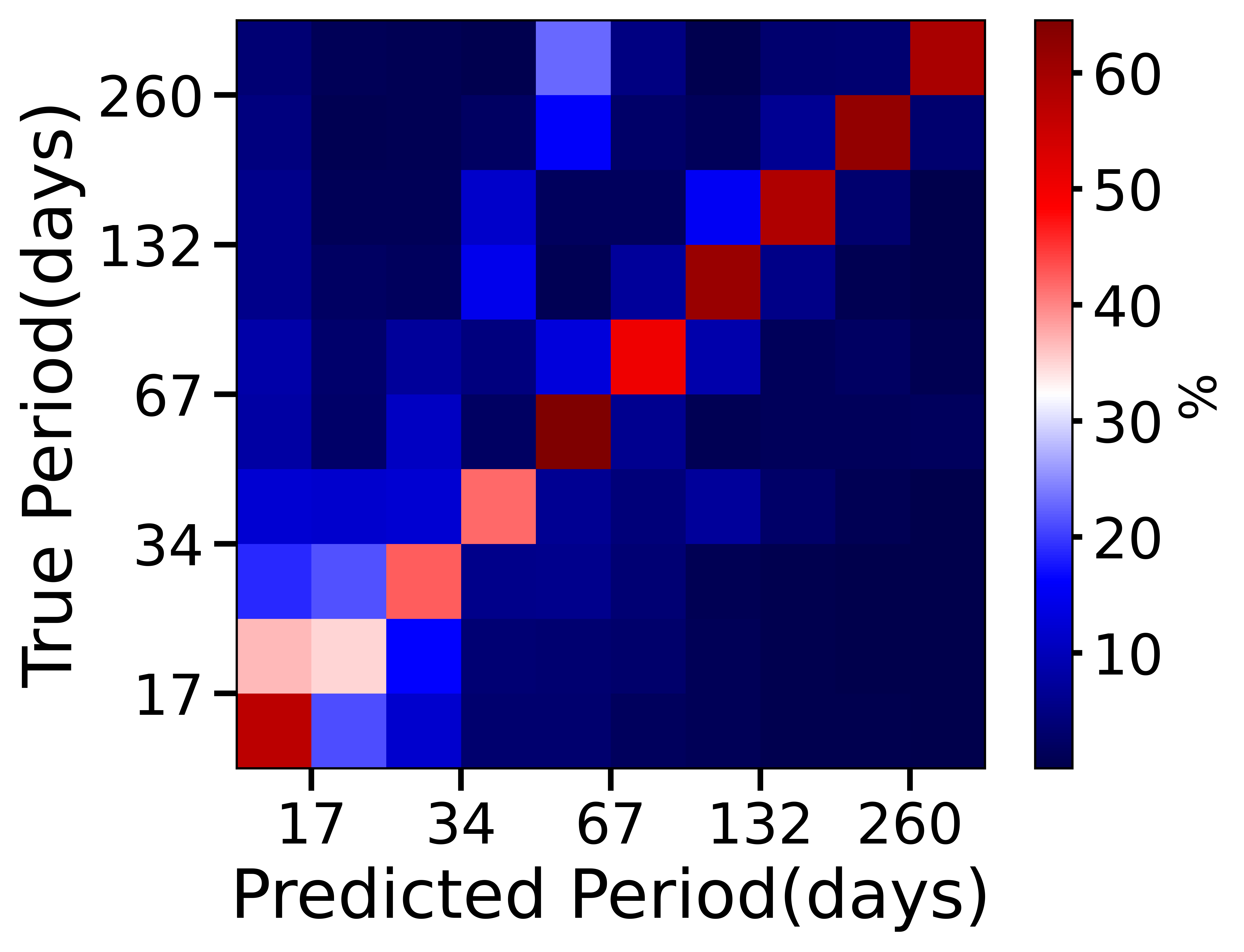}
    \caption{This confusion matrix depicts the model's performance in predicting orbital periods for ordered dataset V3 after fine-tuning. The matrix is normalized for each bin along the \enquote{True Period} axis. The central line, for periods above approximately 35 days, shows high model accuracy for that range. Below this threshold, solar rotation significantly impacts period prediction accuracy, even after fine-tuning, as seen in the plot.}  

    \label{fig:my_lb17}
\end{figure}

\begin{figure}
    \center
    \includegraphics[width=\columnwidth]{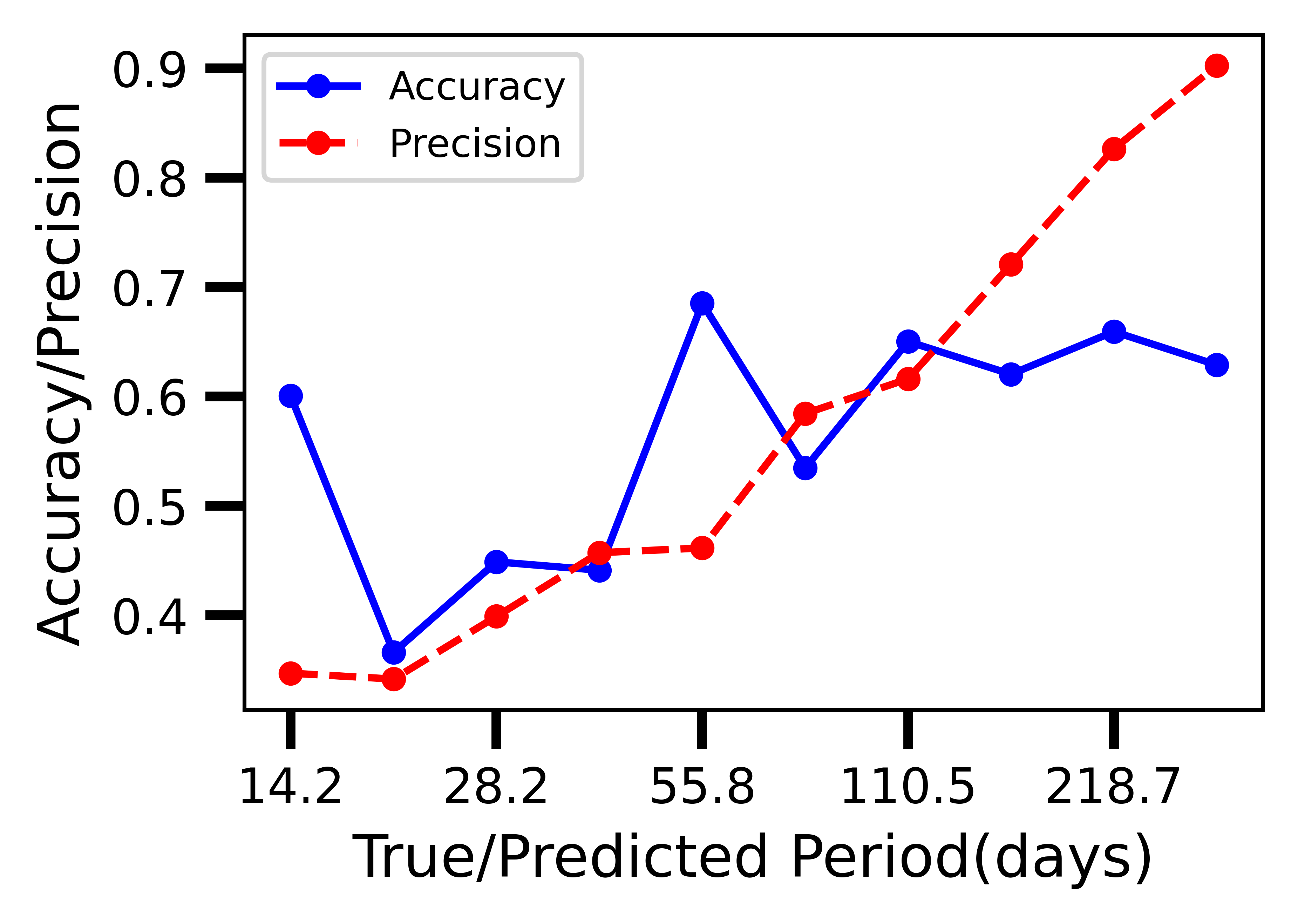}
    \caption{For the ordered dataset V3, accuracy and precision (as described previously in Figure \ref{fig:my_label12}) offer complementary insights into the model's performance in predicting orbital periods. Accuracy declines near values corresponding to the solar rotation rate, suggesting that the rotational signal retains some ambiguity despite fine-tuning, resulting in frequent misclassifications around this period. In contrast, precision increases monotonically with the orbital period. This upward trend reflects a systematic bias where misclassifications are skewed toward lower period values. Consequently, high-period predictions are less likely to be incorrectly assigned to shorter periods, leading to improved precision at longer orbital periods. This pattern suggests that while the model struggles to differentiate signals near the stellar rotation period, it demonstrates greater confidence and reliability in its high-period classifications.
    }
    \label{fig:my_lb18}
\end{figure}

\begin{figure*}
    \center
    \includegraphics{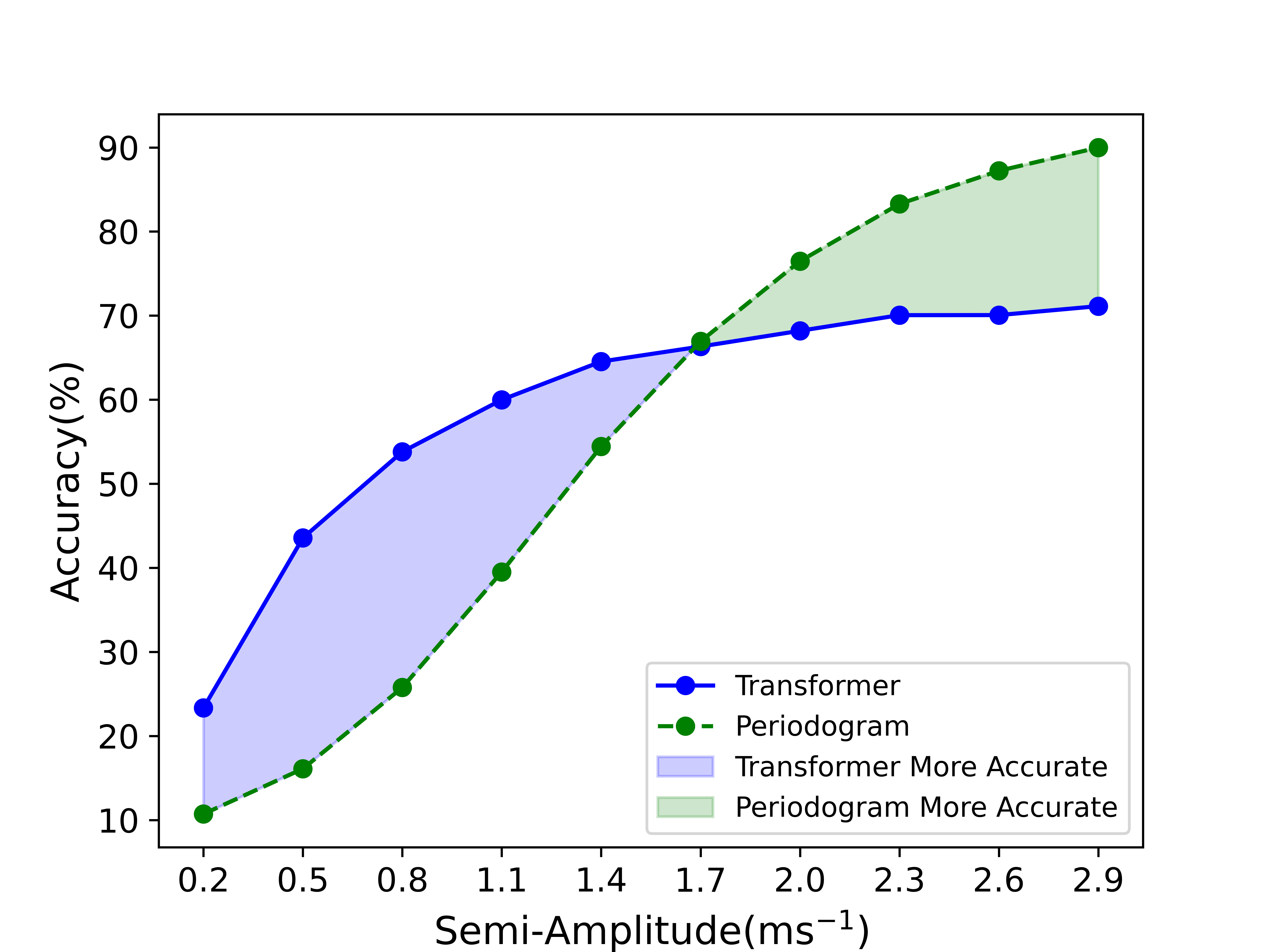}
    \caption{This figure presents a comparison of accuracy between the Lomb-Scargle periodogram and our machine learning model for classifying orbital periods for the ordered dataset V3, using the same discretization as in Figure \ref{fig:my_lb11}. The periodogram achieves higher accuracy at high amplitudes (approximately 1.7 ms$^{-1}$), whereas the model demonstrates superior performance at low amplitudes, outperforming the periodogram by a factor of about 2.}
    \label{fig:my_label19}
\end{figure*}

\begin{figure}
    \center
    \includegraphics[width=\columnwidth]{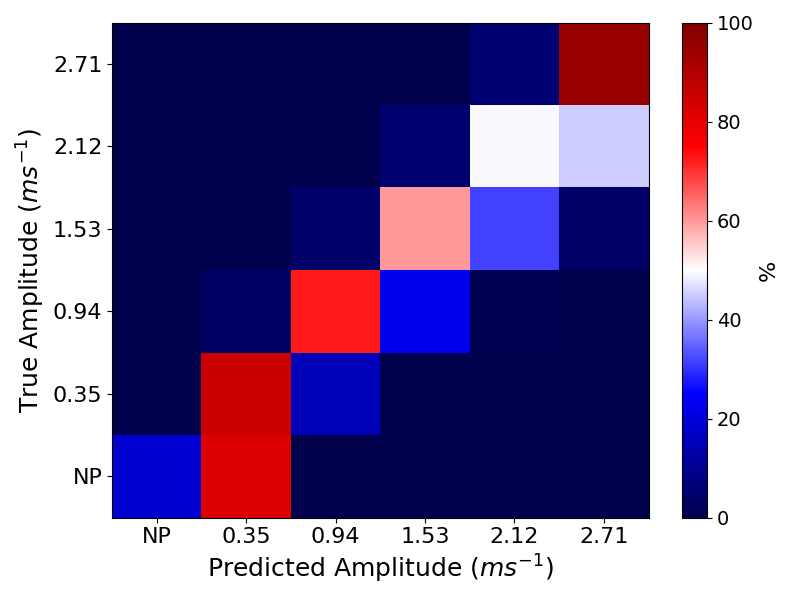}
    \caption{
    This confusion matrix illustrates the model's performance in predicting semi-amplitudes on the unshuffled dataset V3, post-finetuning. The matrix is normalized along the \enquote{True Semi-Amplitude} axis for each bin, and includes the \enquote{No Planet} (NP) scenario. 
    }
    \label{fig:amp_ftune}
\end{figure}

\begin{figure}
    \centering
    \includegraphics[width = \columnwidth]{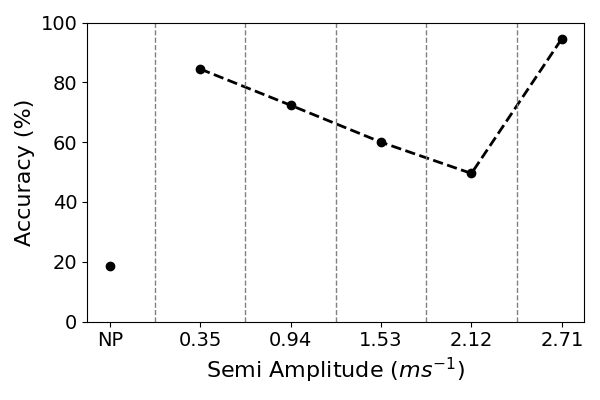}
    \caption {
    This figure shows the classification accuracy for semi-amplitude predictions in the unshuffled dataset V3, post-finetuning. The model employs a six-class scheme comprising five linearly spaced bins representing increasing planetary semi-amplitudes, along with a separate \enquote{No Planet} (NP) category. It achieves an overall accuracy of 74\% across all planetary systems. Accuracy is highest for the lowest and highest amplitude bins and steadily declines across the intermediate bins. The NP category shows significantly lower accuracy, with many cases misclassified into the lowest amplitude bin.
    }  

    \label{fig:amp_ftune_acc}
\end{figure}

\begin{figure}
    \center
    \includegraphics[width = \columnwidth]{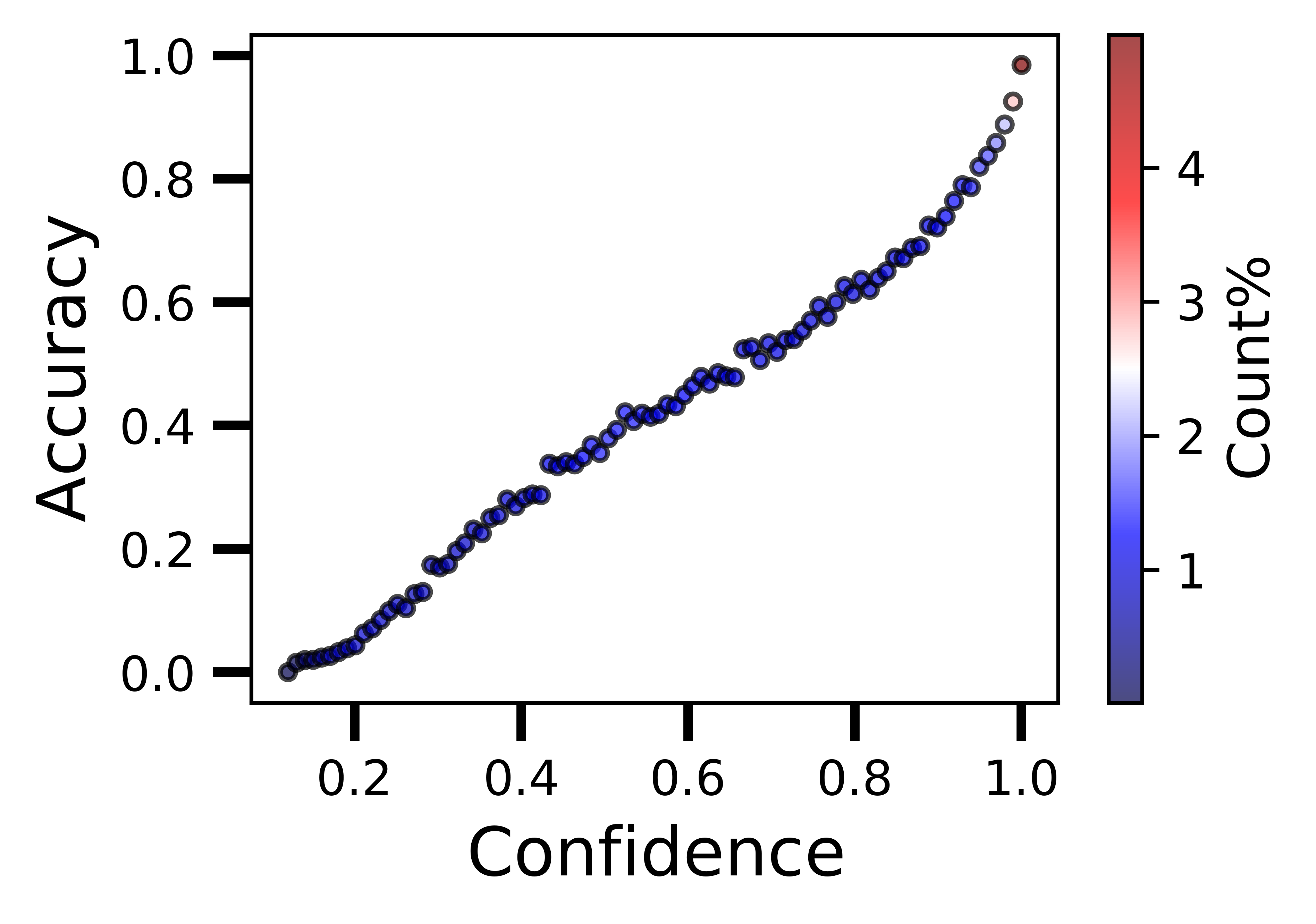}
    \caption{This plot illustrates the relationship between confidence and accuracy for period predictions on the ordered dataset V3. Similar to the previous analysis in Figure \ref{fig:my_lb13}, the 0–1 confidence range is divided into 100 bins, with accuracy values plotted for each. Accuracy improves as the model's confidence in its predicted orbital periods increases. Low-confidence predictions exhibit poor accuracy, while high-confidence predictions converge to 1. However, for the ordered dataset, confidence values never reach unity.}
    \label{fig:my_label20}
\end{figure}

\begin{figure}
    \center
    \includegraphics[width = \columnwidth]{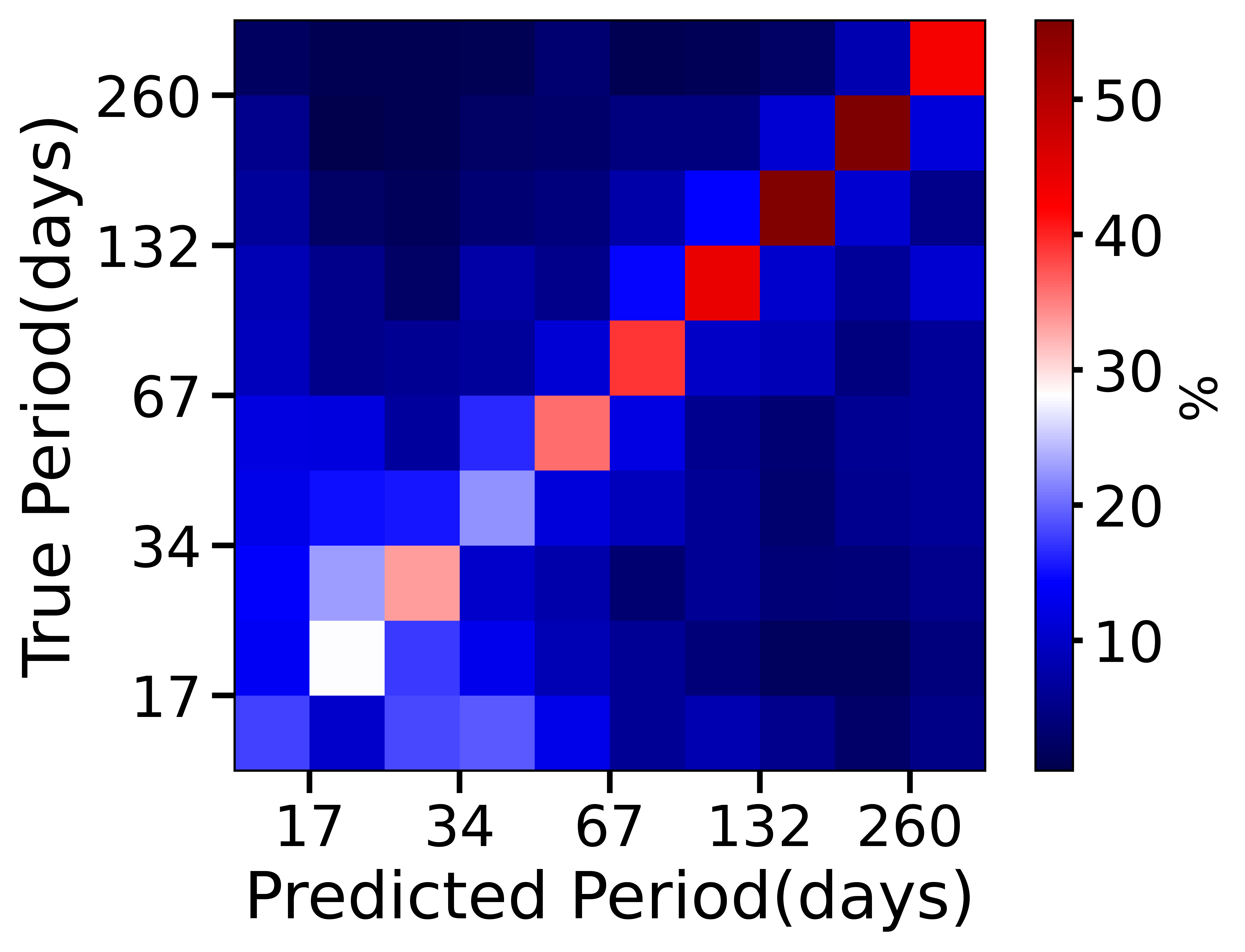}
    \caption{This figure shows the confusion matrix for period classification on the ordered, monthly separated validation dataset M without fine-tuning, normalized along the \enquote{True Period} axis. Unlike Figure \ref{fig:my_lb16}, the model does not show a strong bias toward the stellar rotation rate, with accuracy improving for longer periods.}
    \label{fig:my_lb21}
\end{figure}

\begin{figure}
    \center
    \includegraphics[width = \columnwidth]{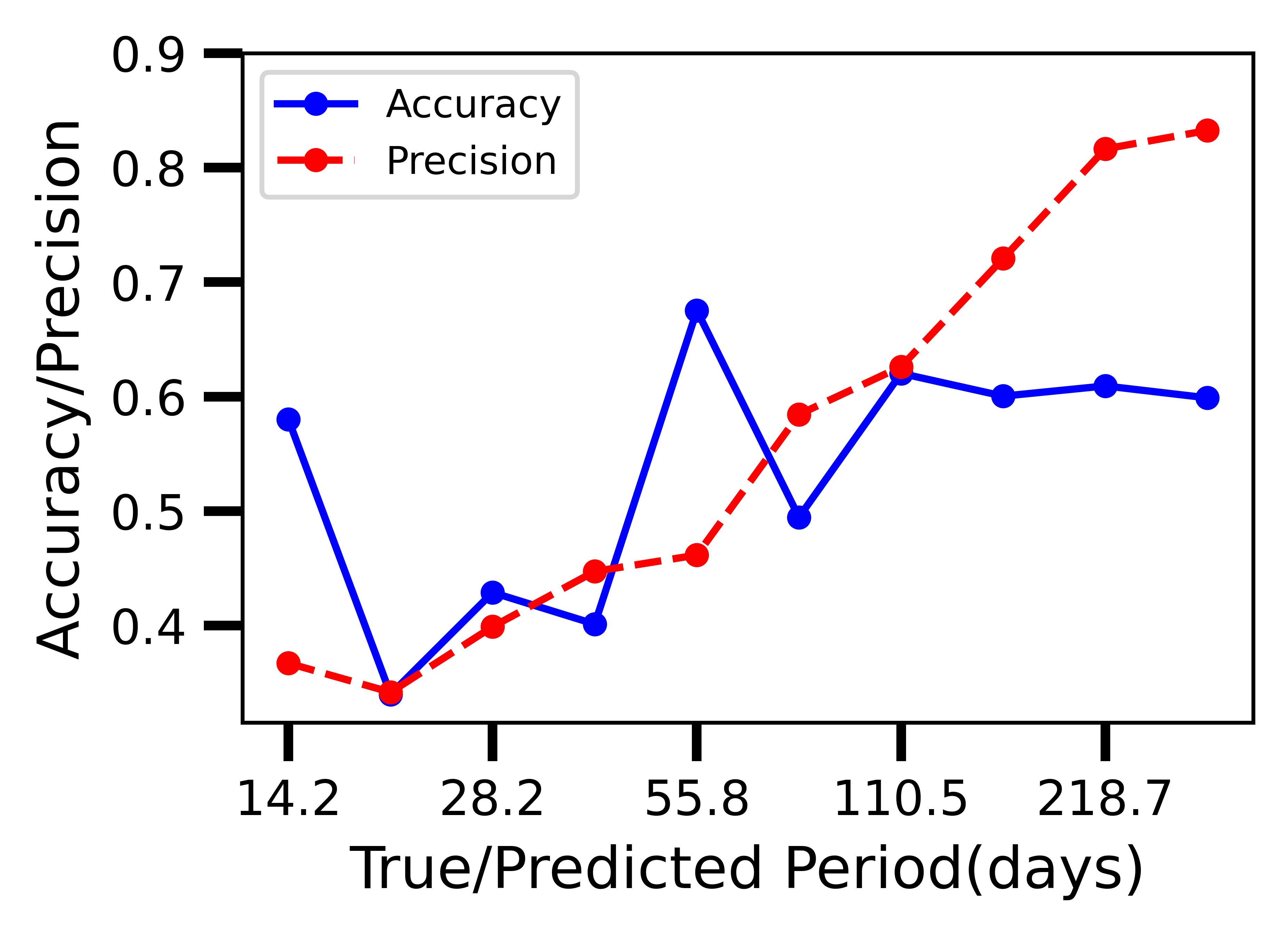}
    \caption{A similar pattern to Figure \ref{fig:my_lb18} is observed in the ordered dataset M with monthly separation. Accuracy declines near the solar rotation rate due to residual ambiguity despite fine-tuning, while precision increases with orbital period as misclassifications are biased toward lower values. This trend enhances precision at higher periods but results in the underprediction of some true values.}
    \label{fig:my_lb22}
\end{figure}

\begin{figure}
    \center
    \includegraphics[width = \columnwidth]{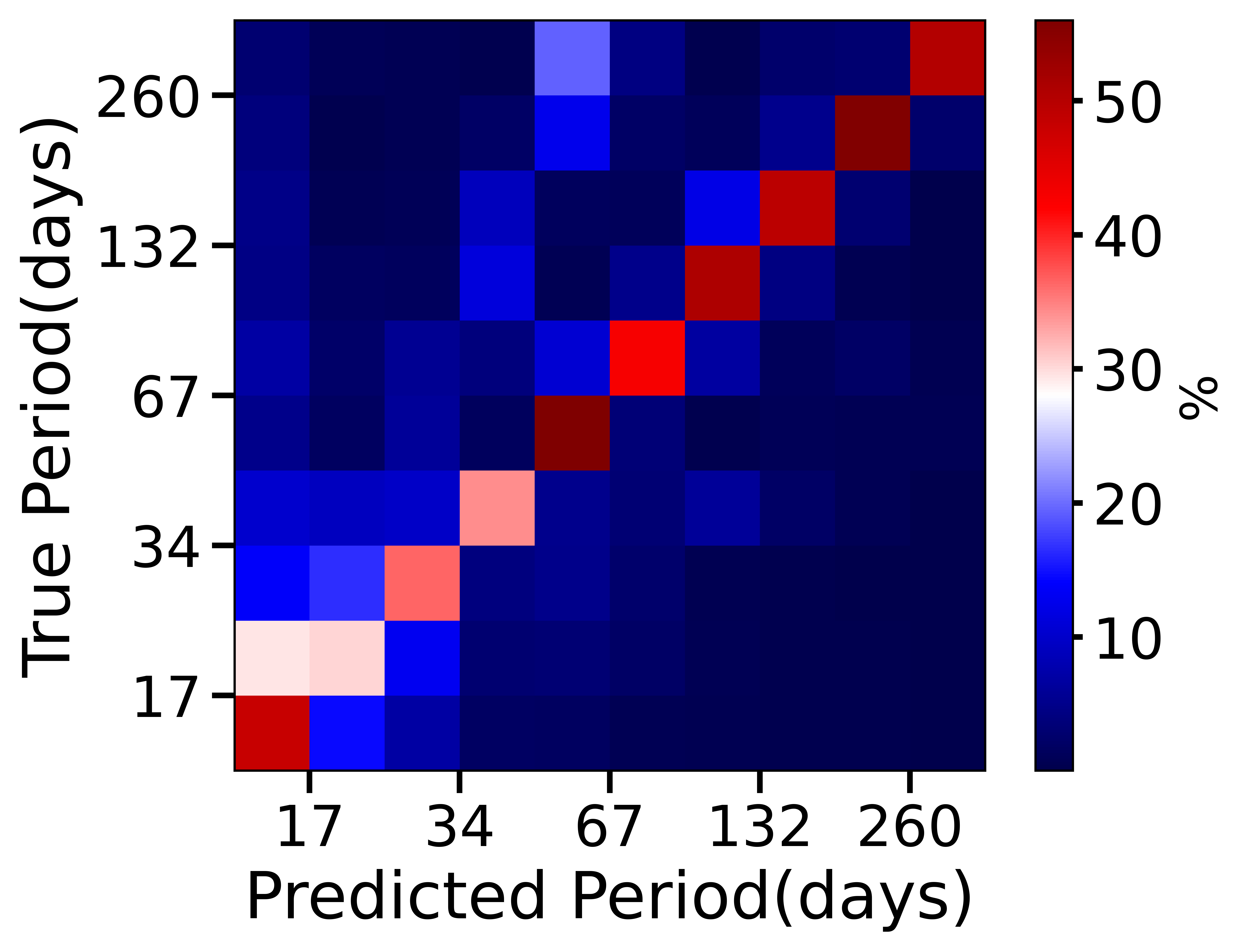}
    \caption{This figure presents the confusion matrix for period predictions on the ordered, monthly separated validation dataset M, after fine-tuning, normalized along the \enquote{True Period} axis. The distribution closely resembles that of Figure \ref{fig:my_lb17}, with the influence of solar rotation remaining apparent for periods shorter than 35 days.}
    \label{fig:my_label23}
\end{figure}

\begin{figure*}
    \center
    \includegraphics{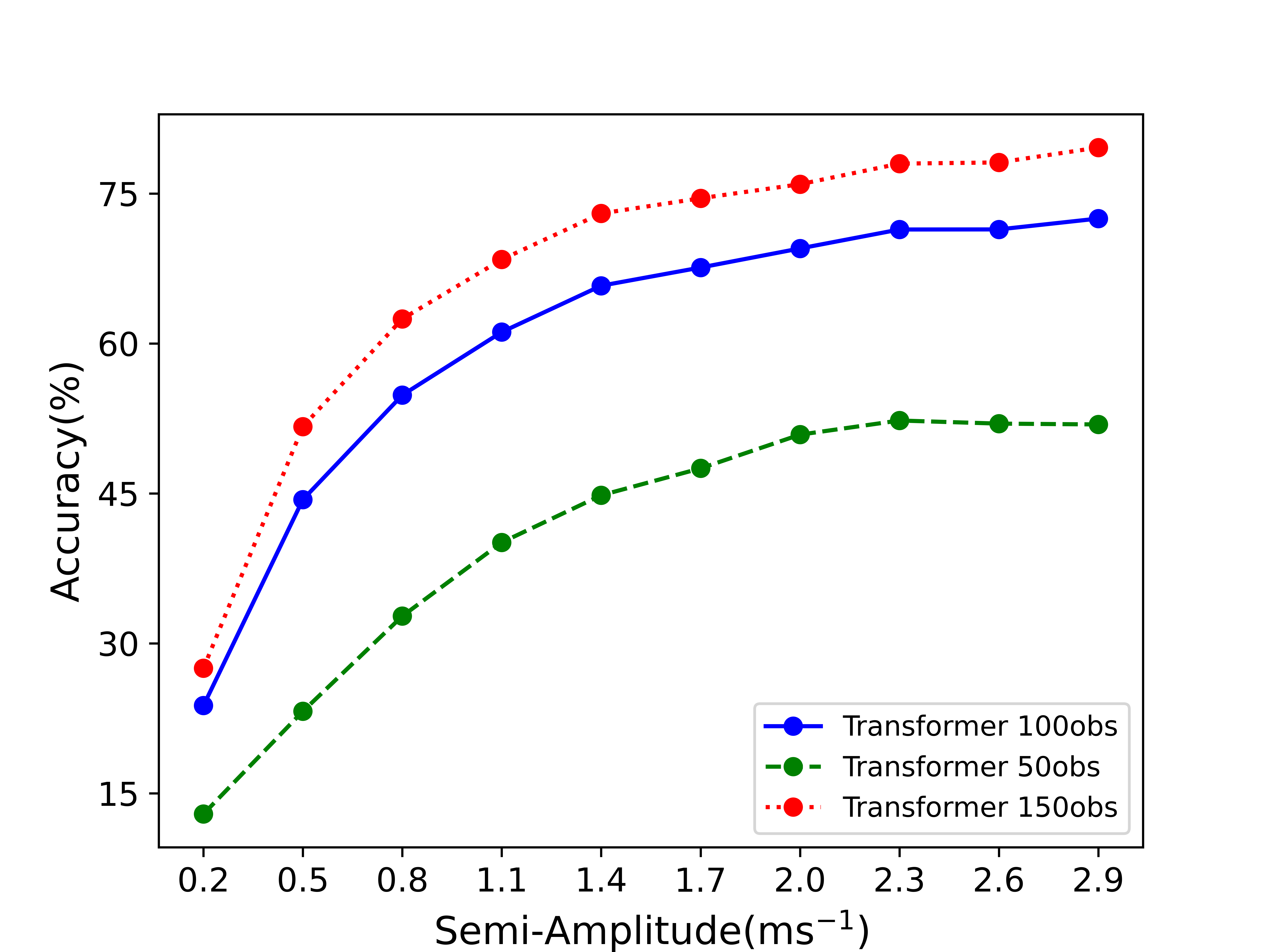}
    \caption{This figure compares the model's classification accuracy for the monthly separated ordered validation dataset M (see Sections \ref{subsec:4.2}, \ref{subsec:5.3}) across scenarios with 50, 100, and 150 observations. Accuracy consistently improves across all semi-amplitudes as the number of observations per sample increases, reflecting the expected effect of a larger observation count.
    }
    \label{fig:my_label_comp}
\end{figure*}

\begin{figure*}
    \center
    \includegraphics[scale = 0.65]{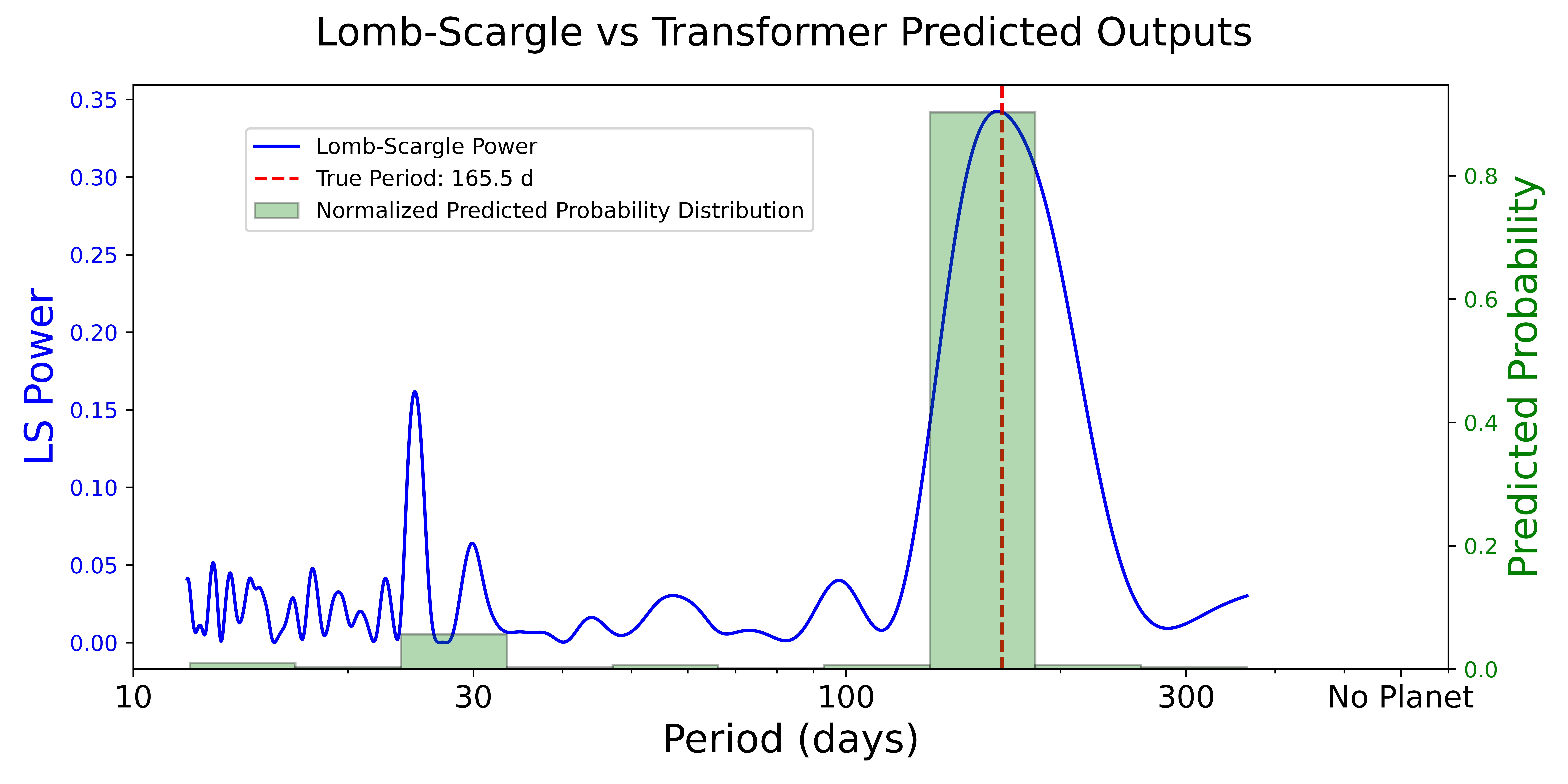}
    \caption{This figure shows a comparison between the Lomb–Scargle periodogram and the model-predicted probability distribution for a representative Sun–planet system. Both methods identify the planetary period to be approximately 165 days. A secondary peak, likely associated with stellar rotation near 25 days, is visible in the periodogram power spectrum. The machine learning model outputs a discrete probability distribution across 11 classes, representing 10 period bins and one class for the no-planet scenario. In contrast, the periodogram provides a continuous power distribution over periods ranging from 12 to 365 days. The x-axis is plotted on a logarithmic scale to better visualize the broad range of periods. For comparison purposes, the period bin corresponding to the highest periodogram power (165 days in this example) is taken as the periodogram-predicted bin, as described in Section~\ref{subsec:5.5}.
    }
    \label{fig:sample_output}
\end{figure*}

\subsubsection{Finetuning}

\label{subsec:5.2.1}

Fine-tuning is performed on a model initially trained on shuffled data, enabling it to adapt to the temporal dependencies of ordered datasets while retaining its previously learned features. This process involves constructing ordered training and validation sets while maintaining the dataset split described in Section \ref{subsec:4.1}. 

Unlike shuffled data, ordered datasets preserve both temporal structure and relative timestamps, allowing the model to refine its ability to distinguish planetary signals from stellar rotation more effectively.

The pre-trained model, initially trained on shuffled data, is fine-tuned on the ordered dataset using a reduced learning rate and a limited number of epochs. This process allows the model to adapt to sequential structures while preserving previously learned features.

\subsubsection{Orbital Period}
\label{subsec:5.2.2}

Fine-tuning significantly decorrelates the Keplerian orbital period from the solar rotation period of 25 days, especially for orbital periods $\gtrapprox$ 35 days. However, accuracy declines at shorter periods, with increased misclassification, and predictions are often influenced by the solar rotation signal. 

Analogous to the drop in accuracy observed in the shuffled dataset (see Section \ref{subsec:5.1.1}), removing the activity-sensitive spectral lines from the CCCF representation leads to a modest accuracy decline of about 3\%, implying that these features retain relevance even in the fine-tuning procedure.

In cases where no Keplerian signal is present, the model continues to predict a spurious period instead of identifying the absence of a planetary companion. This behavior contrasts with the shuffled dataset result shown in Figure \ref{fig:my_label12}, and is discussed further in Appendix \ref{appendix:NoplanetcaseP}. While the model effectively recovers Keplerian periods even in the presence of stellar variability, it does not reliably reject non-planetary signals.

Figure \ref{fig:my_lb17} shows the corresponding confusion matrix for the period predictions on all samples that contain planetary signals. Figure \ref{fig:my_lb18} illustrates how prediction accuracy varies with orbital period. Figure \ref{fig:my_label19} shows how prediction accuracy varies with semi-amplitude, while Figure \ref{fig:my_label20} depicts the relationship between accuracy and confidence scores.

\subsubsection{Semi-amplitude}
\label{subsec:5.2.3}

Semi-amplitude predictions are less affected by temporal ordering than period predictions. However, when the model trained on shuffled data is applied to the ordered V3 dataset, accuracy decreases by 25\% compared to V1. Fine-tuning mitigates this decline, improving accuracy by 22\%. The resulting confusion matrix is presented in Figure~\ref{fig:amp_ftune}, and the corresponding accuracy trend across datasets is shown in Figure~\ref{fig:amp_ftune_acc}. The treatment of the \enquote{No Planet} scenario in this setting is discussed in detail in the Appendix~\ref{appendix:NoplanetcaseK}.

\subsection{For Monthly Separated Data}
\label{subsec:5.3}

We applied a similar methodology to the monthly separated dataset, first training the model on shuffled data, followed by fine-tuning on ordered data. The shuffled model's performance differed considerably from that of set V3 (see Figures \ref{fig:my_lb16}, \ref{fig:my_lb21}), demonstrating reduced accuracy at shorter period values and improved accuracy at longer period values.

After fine-tuning, the prediction accuracy and precision of the ordered validation dataset M (Section \ref{subsec:4.2}) closely matched those of set V3. Figure \ref{fig:my_lb22} shows the variations in accuracy and precision for this validation dataset, while Figures \ref{fig:my_lb21} and \ref{fig:my_label23} compare the model's predictions on ordered data before and after fine-tuning. Notably, in contrast to dataset V3, the monthly separated data and its corresponding model do not predict the solar rotation rate in the absence of fine-tuning.

\subsection{Comparison using Different Numbers of Observations}
\label{subsec:5.4}

To assess the impact of the number of observations on period bin prediction accuracy, we applied our algorithm to orbital parameter estimation using 50 and 150 observation scenarios as well. As expected, accuracy improves with an increasing number of observations. Figure \ref{fig:my_label_comp} illustrates this trend, showing how prediction accuracy varies across fine-tuned validation datasets for these different observation scenarios.

\subsection{The Periodogram Comparison}
\label{subsec:5.5}

We compare our period predictions with those derived from the traditional Lomb-Scargle periodogram method to evaluate the relative accuracy of the two approaches.

\subsubsection{Procedure}
\label{subsec:5.5.1}

The Lomb-Scargle periodogram produces a power spectrum that estimates the likelihood of periodic signals across a range of periods. Typically, the highest peak in this spectrum corresponds to the most probable period. To facilitate a meaningful comparison with our machine learning model, which discretizes the period range and assigns a label corresponding to the interval in which the period most likely resides, we treat the periodogram peak in a similar manner.

Specifically, we extract the period corresponding to the peak power in the Lomb-Scargle spectrum and assign it to the appropriate period bin, analogous to the classification performed by our model. In doing so, both approaches yield discrete period class predictions, allowing for a direct comparison.

Figure~\ref{fig:sample_output} shows the Lomb-scargle periodogram power distribution and the model probability output for a sample Sun-planet system.

An accurate prediction is defined as the assignment of the maximum power period (from the periodogram) or the predicted bin (from the Transformer) to the bin containing the true period. No threshold is applied, and no penalty is imposed for high false alarm probabilities or low peak amplitudes in the periodogram, or low confidence in the model’s prediction. This allows for a consistent evaluation of performance across methods based on discrete period bins.

\subsubsection{Results of the Comparison}
\label{subsec:5.5.2}

For set V1, our machine learning algorithm demonstrates higher accuracy than the periodogram, particularly in the first three bins where K $\lesssim$ 0.95 ms$^{-1}$. However, beyond this range, the periodogram shows slightly higher accuracy than the machine learning predictions, starting from the fourth bin (see Figure \ref{fig:my_lb11}). 

For the temporally ordered set V3, the periodogram achieves lower accuracy than the model at low amplitudes but improves as amplitude increases (see Figure \ref{fig:my_label19}). It surpasses the model's accuracy at approximately 1.7 ms$^{-1}$. Beyond this threshold, the model exhibits signs of overfitting, with training accuracy continuing to improve while validation accuracy plateaus. This performance plateau is likely due to the limited size of the ordered dataset, restricting the model's ability to effectively generalize and learn time-dependent patterns.

For high-amplitude ($>$1.5 ms$^{-1}$) period predictions, it is notable that approximately 45\% of the incorrect predictions fall into period bins adjacent to the true value, with slightly lower yet comparable probabilities. Some misclassifications also exhibit a bimodal probability distribution, where the secondary peak aligns with the true period. These findings indicate that even when the model does not predict the exact period, it effectively identifies the surrounding region with high confidence. 

Additionally, these comparative results do not fully incorporate the relative likelihoods of period estimates from both methods. 

A comprehensive summary of the results for predicting orbital periods is presented in Table \ref{tab:2}.

\begin{table}[h]
    \centering
    \caption{Summary of Results for Period Prediction}
    \label{tab:2}
    \begin{tabular}{cccc}
        \toprule
        \textbf{Dataset} & \textbf{Timestamps} & \textbf{Finetuned} & \textbf{Accuracy}  \\ 
        \midrule
        V1 & Shuffled & No & 86\% \\  
        V2 & Ordered, Scaled & No & 39\% \\  
        V3 & Ordered & No & 19\% \\   
        V3 & Ordered & Yes & 54\% \\  
        M & Ordered & No & 33\% \\ 
        M & Ordered & Yes & 55\% \\ 
        \bottomrule
    \end{tabular}
\end{table}

\section{Discussion}
\label{sec:6}


Our results demonstrate that a machine learning approach can outperform standard periodogram methods in the early detection of planetary candidates, particularly for sub-ms$^{-1}$ semi-amplitude signals in solar radial velocity data. Figure \ref{fig:my_lb11} illustrates this for a shuffled dataset with 100 irregularly sampled data points. The improvement is especially pronounced at lower amplitudes, where radial velocity scatter dominates over the Keplerian signal.

A similar improvement in accuracy is observed in the ordered dataset, demonstrating the model's effectiveness in accounting for time-correlated noise, which enhances its predictive performance. Models trained on time-separated data (see Section \ref{subsec:5.2}, Figure \ref{fig:my_lb17}) and monthly-separated data (see Section \ref{subsec:5.3}, Figure \ref{fig:my_label23}) perform similarly on their respective datasets. However, this consistency does not carry over to shuffled data (see Figures \ref{fig:my_lb16}, \ref{fig:my_lb21}), where the choice of training set significantly influences the results.

Specifically, a model trained on fully time-separated data tends to predict the Sun's rotation period when applied to time-ordered data (see Figure \ref{fig:my_lb16}). In contrast, a model trained on monthly-separated data better classifies higher orbital periods, though shorter periods remain more challenging to resolve (see Figure \ref{fig:my_lb21}). 

 By effectively isolating periods of interest, particularly in low-amplitude regimes, our approach offers a robust alternative for detecting planetary signals in noisy radial velocity data.

\subsection{The Aperiodicity Problem}
\label{subsec:6.1}

Standard machine learning models, such as CNNs\citep{CNN_Lecun1998} and LSTMs\citep{LSTM_Hochreiter1997} (see Appendix \ref{appendix:CNNLSTM}), perform well when predicting orbital parameters from regularly sampled data (i.e., equal time intervals). However, their accuracy deteriorates when dealing with aperiodic timestamps. This limitation is particularly relevant in astrophysical applications, where observations are often irregularly sampled due to various observational constraints.

Consequently, machine learning models capable of processing irregularly sampled data are essential for accurately characterizing real astrophysical observations.

The Vision Transformer (ViT) we use in this study offers a compelling alternative for handling aperiodic timestamps. Unlike CNNs and LSTMs, which struggle with irregular time intervals, ViT's attention mechanism effectively processes non-uniformly sampled data. This capability makes it particularly well-suited for analyzing and predicting astrophysical phenomena based on real observational datasets \citep{dosovitskiy2021image}.

\subsection{Limitations}
\label{subsec:6.2}

A major limitation in improving these models is the scarcity of large, uniformly processed datasets, which can be critical for robust machine learning applications. 

The Sun remains the only star with an extensive radial velocity (RV) dataset, enabling machine learning models to be trained directly on its observations without requiring external priors. In contrast, applying similar methods to other stars necessitates transfer learning, as their RV datasets are significantly smaller. 

Our current analysis is based on 19 months of NEID solar observations, a dataset that will expand over time. However, with the present dataset size, the number of independent samples is insufficient to capture and generalize long-term, time-correlated stellar activity signals comprehensively.  As the dataset expands in future model iterations, it will enable a more detailed characterization and mitigation of activity-driven variations, thereby enhancing the model’s ability to distinguish planetary signals from stellar noise.

The current model architecture requires a fixed number of observations for each training instance. For a star with $N$ observations, the model must be trained specifically for that $N$, necessitating a complete retraining process. Training on 840,000 samples and validating on 500,000 samples using a GPU (NVIDIA RTX A4000, 15.35 GB memory, CUDA version 12.4) requires approximately 46 hours. Despite this computational cost, the model remains adaptable, as it can be efficiently retrained for different N, making it versatile for various observational datasets.

Extending the model to handle variable-length time series is theoretically possible. However, training and evaluating such data introduce additional challenges, including inconsistencies in temporal structures and the need for specialized architectures capable of dynamically processing sequences of varying lengths. These complexities necessitate alternative approaches, which are beyond the scope of the current implementation.

\subsection{Applications}
\label{subsec:6.3}

The accuracy of correctly identifying the period bin obtained with just 100 observations over approximately one year highlights the effectiveness of this method in identifying planetary candidates around solar-type stars. Unlike periodogram tests during an RV survey that require long baselines and a large number of observations to identify potential planetary candidates, our approach achieves comparable results with significantly shorter observation baselines. This underscores the advantage of ML-based approaches, which can efficiently extract potential planetary signals, even from shorter datasets.

Candidates identified through this method can be prioritized for targeted follow-up observations, with subsequent data collection optimized to refine period estimates and confirm planetary signals. The validation sets V3 and M reveal instances where the periodogram exhibits low likelihood power for semi-amplitudes below 1.5 ms$^{-1}$, while the ML model confidently predicts a strong signal. In these cases, when the periodogram makes an incorrect prediction and the ML model is correct, the model typically demonstrates moderately high confidence ($>$ 0.7) in approximately 51\% of such instances. These instances emphasize the model’s ability to uncover promising planetary candidates that may otherwise be overlooked, demonstrating its potential as a complementary tool in RV planet searches.

\subsection{Next Steps in Development}
\label{subsec:6.4}

The current model does not yet represent the upper limit of achievable performance with existing resources, and several refinements can further enhance its accuracy.

One potential improvement involves incorporating separately averaged cross-correlation functions (CCFs) for red and blue spectral lines. Since these spectral regions contain different astrophysical information, leveraging their distinct properties may improve the model’s predictive capabilities  \citep{Dumusque_2018}.

Future studies will evaluate the model’s applicability to other G-type stars and test datasets to assess its generalizability across diverse stellar populations and instrument configurations. We also plan to improve our finetuning step using simulated time-correlated spectral data from simulations like SOAP-GPU \citep{SoapGPUrefId0, SoapGPUrefId1} and StarSim\footnote{\label{Starsim}\href{https://github.com/dbarochlopez/starsim}{StarSim}} \citep{Starsimref} without unlearning the training on the real solar data.

Apart from the application of our model in detecting weak signals, we plan to apply this framework to classify strong signals as well. Very often, stars show strong statistically significant peaks in periodograms, and one has to rule out whether this is a false positive due to stellar activity (instead of a real planet signal). In our follow-up model, we address this problem as a binary classification challenge for a Transformer model.

Building on these efforts, our longer-term goal is to consolidate these approaches into a unified toolset for the community. We tentatively name this framework ViPer-RV (Vision Transformers for Periodicity in RV analyses), which we envision as a resource for both detecting subtle planetary signals and classifying periodicities in radial velocity data based on their origin.

\section{Conclusion}
\label{sec:7}

The Earth-Sun system, with its one-year period and RV amplitude of 9 cms$^{-1}$, serves as an example of a long-period, low-amplitude planetary system. Detecting such systems requires robust mitigation of stellar activity due to their long periods and weak signals.  Traditional methods selectively use activity-sensitive spectral regions, limiting their ability to fully exploit all available spectral information.

Our machine learning approach is designed to maximize the use of available spectral data to identify and isolate periodic signals and potentially be generalizable across different instruments. By analyzing subtle variations in spectral line shapes and shifts, the model differentiates stellar RV variability from Keplerian motion.

The model is trained and tested on effectively 100-epoch solar observations injected with Keplerian signals (post-barycentric correction). For shuffled datasets, this approach achieves a validation accuracy of 86\% for orbital period predictions and 76\% accuracy for semi-amplitude predictions.

In the ordered NEID solar dataset, where temporal correlations in stellar activity are preserved, the model's accuracy decreases to $\approx$ 54\%. Despite this decline, it continues to outperform the Lomb-Scargle periodogram at low amplitudes ($<1$ms$^{-1}$) by approximately a factor of two. As semi-amplitude increases, the periodogram's performance improves, eventually surpassing the model's accuracy at $\approx$1.7 ms$^{-1}$.

At high amplitudes ($>$ 1.5 ms$^{-1}$), around 50\% of incorrect predictions are either in adjacent bin values to the true period or display bimodal distributions, with the secondary peak corresponding to the true period value. This behavior suggests that even when the model does not precisely recover the orbital period, it consistently identifies the correct region in parameter space, reinforcing its reliability in detecting planetary candidate signals within stellar RV datasets.

In short, our findings demonstrate that machine learning can enhance the extraction of planetary signals from radial velocity data, particularly in the low semi-amplitude regime ($<$1 ms$^{-1}$) where traditional periodogram-based methods struggle. This approach improves the efficiency of RV surveys by enabling more robust detections of low-mass exoplanets and refining candidate selection for follow-up observations.

While applied here to solar data, this framework is adaptable to stellar RV datasets, making it a promising tool for ongoing and future exoplanet searches. Further improvements will focus on distinguishing true planetary signals from stellar activity-induced variations, a key challenge in high-precision RV measurements.

As spectrographs push toward 10 cms$^{-1}$ precision, machine learning techniques will be crucial in mitigating stellar noise and maximizing the scientific yield of next-generation RV surveys.

\section*{acknowledgements}

We thank the referee for the detailed feedback and suggestions that improved the clarity and content of the manuscript. We also acknowledge support from the Department of Atomic Energy, Government of India, under Project Identification No. RTI 4002. This research was supported in part by a generous donation from the Murty Trust, an initiative of the Murty Foundation, aimed at enabling advances in astrophysics through the use of machine learning. The Murty Trust is a not-for-profit organization dedicated to the preservation and celebration of culture, science, and knowledge systems born out of India, headed by Mrs. Sudha Murty and Mr. Rohan Murty.

We would like to thank Professor Suvrath Mahadevan, Professor Eric Ford, Dr. Paul Robertson, Mr. Siddharth Dhanpal, Mr. Prasad Subramanian, and Mr. Nipun Ghanghas for their insightful discussions and valuable advice. Their contributions have been instrumental in shaping this work.

We thank Professor Xavier Dumusque for insightful feedback, particularly regarding the inclusion and interpretation of the \enquote{No Planet} scenario. His suggestion helped clarify that while the model tends to assign spurious periods to no-planet cases, typically below 45 days, predictions at longer periods are less affected, indicating that the $\gtrapprox$ 45-day regime is comparatively robust against contamination from non-planet scenarios. A similar observation was also seen in Semi-Amplitude predictions (see Appendix \ref{appendix:Noplanetcase} for both observations).

This paper contains data taken with the NEID instrument, which was funded by the NASA-NSF Exoplanet Observational Research (NN-EXPLORE) partnership and built by Pennsylvania State University. NEID is installed on the WIYN telescope, which is operated by the National Optical Astronomy Observatory, and the NEID archive is operated by the NASA Exoplanet Science Institute at the California Institute of Technology. NN-EXPLORE is managed by the Jet Propulsion Laboratory, California Institute of Technology under contract with the National Aeronautics and Space Administration.

We also acknowledge the use of ChatGPT, developed by OpenAI, for assistance in language editing during the preparation of this manuscript.

\facility{WIYN (NEID)}

\software{
\texttt{TensorFlow} \citep{tensorflow_abadi2016tensorflowlargescalemachinelearning}, \texttt{pytorch} \citep{torch_paszke2019pytorch}, \texttt{NEID DRP} \citep{neid2023}, 
\texttt{astroquery} \citep{astroquery_Ginsburg_2019}, \texttt{astropy}
\citep{Astropy_Price-Whelan_2018},
\texttt{barycorrpy} \citep{Kanodia_2018}, \texttt{celerite} \citep{celerite_Foreman-Mackey_2017}, \texttt{ChatGPT} \citep{ChatGPT}, \texttt{matplotlib} \citep{matplotlib_4160265}, \texttt{multiprocessing} \citep{python_multiprocessing},   \texttt{numpy} \citep{numpy_5725236},  \texttt{pandas} \citep{pandas2020},  \texttt{radvel} \citep{Fulton_2018}, \texttt{RVEstimator} \citep{RVEstimator}, \texttt{scipy} \citep{scipy}
}

\appendix

\section{Machine Learning Procedure}

Machine Learning algorithms iteratively adjust their internal parameters by learning from labeled input-output pairs in the training data. During this training process, the model identifies underlying patterns by minimizing a loss function, which quantifies the discrepancy between predicted and true outputs. A successful training procedure is characterized by a steady decline in the loss function value as the model improves its input-output mapping.

Once trained, the ML model applies this learned representation to make predictions or classifications on previously unseen data, effectively generalizing beyond the training set.


\subsection{Dataset Splitting}

The final dataset consists of 35,757 1D-CCCF observation vectors obtained after the pruning and processing steps described in Section \ref{sec:3}. To implement our machine learning model, these samples are partitioned into two temporally separated subsets: training and validation datasets.

Each subset is independently processed to generate corresponding 2D-CCCF vectors, as outlined in Section \ref{sec:new4}. These processed datasets are referred to as the training and validation raw datasets.

The training set is used to optimize the model’s internal parameters through iterative updates. The validation set, containing previously unseen samples, is employed to monitor the model's predictive performance and assess overfitting; a situation where the model fits the training data too closely, limiting its ability to generalize to new observations.


\subsection{Training Methodology}

As described in Section \ref{sec:new4}, the training input is a 2D array of 99 rows (derived from 100 original 1D-CCCFs by taking differences relative to the first), each row representing a single observation. The model is trained to map these inputs to probability arrays corresponding to orbital parameters, specifically the orbital period and semi-amplitude. The training dataset, based on the processed observation 1D-CCCF vectors discussed in Section \ref{sec:3}, includes 26,777 such 1D-CCCF samples, with an additional 6,949 samples reserved for the validation set V1 (see Figure \ref{fig:my_label8}). 

The orbital period spans 12 to 365 days, divided logarithmically into 10 bins labeled 0 to 9 for classification. Similarly, the semi-amplitude of the Keplerian signal ranges from 0.05 to 3 ms$^{-1}$, partitioned into 5 equal bins labeled 0 to 4.

The model's objective is to classify the orbital period and semi-amplitude labels based on the structured input samples. These inputs are represented as 99×1722 2D vectors (see Section \ref{sec:new4}).

Throughout training, the model processes the entire dataset iteratively, with periodic evaluations on the validation set to monitor performance and mitigate overfitting. Each complete pass through the training and validation datasets constitutes an epoch. This regular assessment ensures a balance between the model's learning progression and its ability to generalize to unseen data.

This training methodology is integral to our overall framework, where the iterative improvement over multiple epochs allows the model to achieve robust classification accuracy for orbital parameters.

\subsection{Classification versus Regression}

\label{appendix:classvsreg}

In parameter estimation tasks, continuous variables are traditionally predicted using regression models. However, the formulation of the problem significantly influences the performance of machine learning models. Recasting a regression problem as a classification task can often yield improved results \citep{stewart2023regressionclassificationinfluencetask}.

This approach involves discretizing the continuous parameter range into a series of bins. The model is then trained to map input samples to the corresponding bin labels that best represent the target parameter values.
By controlling the number and spacing of these bins, the formulation enables fine-tuning of prediction granularity while balancing against dataset limitations, i.e., trading resolution for stability and tractability where needed.

The output takes the form of a probability vector, indicating the likelihood of the parameter falling within each bin.

We initially explored a regression formulation using MSE loss for period prediction, but observed frequent convergence failures and large errors, particularly for low-SNR signals and cases with overlapping planetary and activity-induced variations. Reformulating the task as classification over discretized period bins significantly stabilized training. This behavior is consistent with theoretical insights and prior findings in similar signal detection tasks \citep{stewart2023regressionclassificationinfluencetask}.

Such classification-based formulations have also been successfully adopted in other areas of astrophysics, for instance, parameter predictions in asteroseismology \citep{siddharth_dhanpal2022ApJ...928..188D}.

One key advantage of this formulation is its robustness to outliers. In regression models, large prediction errors from outliers can disproportionately impact training, leading to unstable results. In contrast, classification models, with their discrete bin structure, reduce this sensitivity by limiting the effect of extreme values \citep{stewart2023regressionclassificationinfluencetask}. This not only improves overall prediction accuracy but also prevents unphysical values outside the defined parameter range.

Additionally, classification can provide a direct measure of prediction uncertainty through the probability distribution across bins, offering clearer insights into the model’s confidence. This probabilistic output is particularly valuable for astrophysical parameter estimation, where uncertainties and predictions are equally critical for robust analysis.

In this work, our primary objective is to focus on identifying regions of interest in the orbital parameter space. The classification output allows us to isolate these regions, which can then be refined with targeted follow-up analysis or higher-resolution modeling in future iterations. This strategy aligns with the broader goal of improving the detection and characterization of planetary signals by progressively narrowing down parameter ranges of interest.

\subsection{ML Architecture}

To predict orbital period and semi-amplitude, we explored multiple machine learning architectures, each evaluated for its ability to distinguish activity-induced RV variations from true Doppler shifts by analyzing structural differences in the cross-correlation function (CCF). We tested four different models: a Convolutional Neural Network (CNN), a Long Short-Term Memory (LSTM) network, a hybrid CNN-LSTM, and a Vision Transformer (ViT).

\subsection{CNN and LSTM}
\label{appendix:CNNLSTM}

A Convolutional Neural Network (CNN) is widely used for image recognition and classification due to its ability to extract hierarchical features from input data \citep{oshea2015introduction}.

CNNs employ convolutional layers that apply learnable filters to the input, generating feature maps that capture structural patterns such as edges, textures, and shapes. These feature maps are then downsampled using pooling layers, which reduce spatial dimensions while retaining critical information. Fully connected layers at the final stage use the extracted features to classify the input.

In contrast, Long Short-Term Memory (LSTM) networks are designed for sequential data processing and excel at capturing long-range dependencies while mitigating issues such as vanishing or exploding gradients \citep{staudemeyer2019understanding}.

LSTMs incorporate memory cells that store information over extended sequences, dynamically updating or discarding information based on relevance. This functionality is controlled by three types of gates:

\begin{itemize}
    \item Input gate: Determines which new information should be stored.
    \item Forget gate: Regulates which stored information should be discarded.
    \item Output gate: Selects relevant information to be passed to the next time step.
\end{itemize}

These mechanisms allow LSTMs to model complex temporal relationships, making them well-suited for time-series analysis, including RV signal prediction. However, both CNNs and LSTMs struggle with irregularly sampled data, limiting their performance on real astrophysical datasets. In our work, we address this challenge by employing a transformer-based architecture, which can inherently handle non-uniform sampling more effectively.

\subsection{Vision Transformer (ViT)}
\label{appendix:ViT}

While CNNs and LSTMs are effective for regularly sampled data, they struggle to handle irregular observational cadences common in astrophysics. The Vision Transformer (ViT) offers a promising alternative by processing non-uniformly sampled data through a self-attention mechanism.

Originally developed for natural language processing (NLP) tasks, the transformer model revolutionized the field by assigning varying importance to different parts of the input data using self-attention \citep{vaswani2023attention}. Unlike traditional sequential models, transformers process input data non-sequentially, capturing both short and long-range dependencies without being constrained by input order.

ViTs adapt this architecture for computer vision by dividing images into sequences of patches, analogous to words in a sentence. This approach exploits the transformer's ability to incorporate global context while retaining local structure, making it well-suited for structured data like spectral time series in RV analysis. By capturing both spectral and temporal dependencies, ViTs can effectively distinguish between stellar activity and planetary-induced Doppler shifts, even with irregular observation timestamps.

In this work, we utilize the ViT architecture to analyze concatenated cross-correlation function (CCCF) data, treating each row as a sequential patch. This approach capitalizes on the model’s ability to process non-uniformly sampled data, providing a compelling solution for time-series analysis in exoplanet detection.

\subsection{Salient Features}

Vision Transformers (ViTs) differ from Convolutional Neural Networks (CNNs) by processing input data as sequences of patches rather than through hierarchical convolutional layers. In typical vision tasks, images are divided into uniform square blocks. In our case, the 2D input matrix is partitioned row-wise, with each row representing the Keplerian signal captured at a distinct time. These patches are then flattened and linearly transformed into contextual embeddings; compact, meaningful representations that preserve essential information while reducing dimensionality \citep{dosovitskiy2021image}.

Contextual embeddings offer two key advantages:

\begin{itemize}
    \item Semantic Representation: They capture meaningful patterns within the data, improving the model's interpretive ability.
    \item Dimensional Reduction: They lower computational complexity, enhancing both model efficiency and training speed.
\end{itemize}

After embedding, the tokens are processed by the transformer's self-attention architecture, aligning with the standard transformer framework.

A notable strength of transformer-based models is their capability for generalization and transfer learning. Once sufficiently trained on a comprehensive dataset, the model can be fine-tuned for similar tasks on different datasets. In principle, a well-generalized ViT trained on solar RV data could be fine-tuned to operate on data from other stars, provided the solar dataset effectively captures the underlying patterns needed for transferability \citep{transferlearning_malpure2021investigatingtransferlearningcapabilities}. 

\subsection{Positional Encoding}

Positional encoding is a fundamental component of Transformer models \citep{vaswani2023attention}, addressing the model's inability to inherently capture positional order. In Vision Transformers (ViTs), where our input spectral representations are processed as a sequence of patches, retaining spatial and temporal information is essential. Positional encoding preserves this information by embedding positional context into the model during training.

Our implementation uses sine and cosine functions to generate the positional encoding vectors. By varying frequencies and phase shifts across dimensions, each position is uniquely represented, ensuring that positional information remains distinguishable throughout the model.

These encoding vectors are added to the patch embeddings before being passed into the Transformer layers. This integration allows the model to simultaneously process spectral information from the patch embeddings and spatial/temporal information from the positional encodings.

In our ViT model, each shifted 1D-CCCF vector is treated as a patch. This design leverages prior knowledge that the spectral and temporal dimensions hold distinct meanings in our 2D \enquote{image} representation, enhancing the model's ability to capture time-dependent patterns.

\subsection{The Self-attention Mechanism}

After positional encoding is applied, the patch embedding process condenses meaningful information from each input patch into a compact vector representation. The self-attention mechanism then captures dependencies and relationships between these patches, enabling the model to interpret contextual interactions within the input data \citep{vaswani2023attention, dosovitskiy2021image}.

This mechanism begins by projecting the embedded vectors into three distinct sets: query (Q), key (K), and value (V) vectors. These projections are trainable parameters optimized during the model's training process.

The Q and K vectors are combined through a dot product operation, producing a square matrix of attention scores. Applying a softmax function normalizes this matrix, converting it into an attention weights matrix. Each row in this matrix represents the attention distribution of a query patch over all key patches.

Multiplying the attention weights matrix by the V vectors yields the final output of the self-attention mechanism, which is then passed to subsequent layers for further processing and eventual parameter estimation. 

This self-attention process is critical for capturing long-range dependencies and contextual relationships across input patches. By dynamically weighting the importance of each patch, the model achieves a nuanced understanding of spectral and temporal information, improving its ability to differentiate between stellar activity and planetary-induced Doppler shifts in RV data.

\subsection{Multi-head attention}

In Transformer models, multi-head attention is implemented to enable the model to learn diverse and complementary attention patterns simultaneously. By distributing the attention mechanism across multiple heads, the model can capture different types of relationships within the input data, improving its capacity to model complex dependencies and accelerating the training process \citep{vaswani2023attention}.

In our ViT model, each attention head processes the input independently, generating distinct outputs that focus on varying aspects of the spectral and temporal information in our RV data. These outputs are then concatenated and passed through a linear projection layer, which integrates information from all heads into a unified representation.

This final linear projection is subsequently mapped to a probability vector representing the orbital parameters, specifically the orbital period and semi-amplitude. By analyzing multiple perspectives simultaneously, the multi-head attention mechanism enhances the model’s predictive performance in distinguishing between stellar activity and planetary-induced Doppler shifts.

\begin{figure*}
    \centering
    \includegraphics[width=\textwidth]{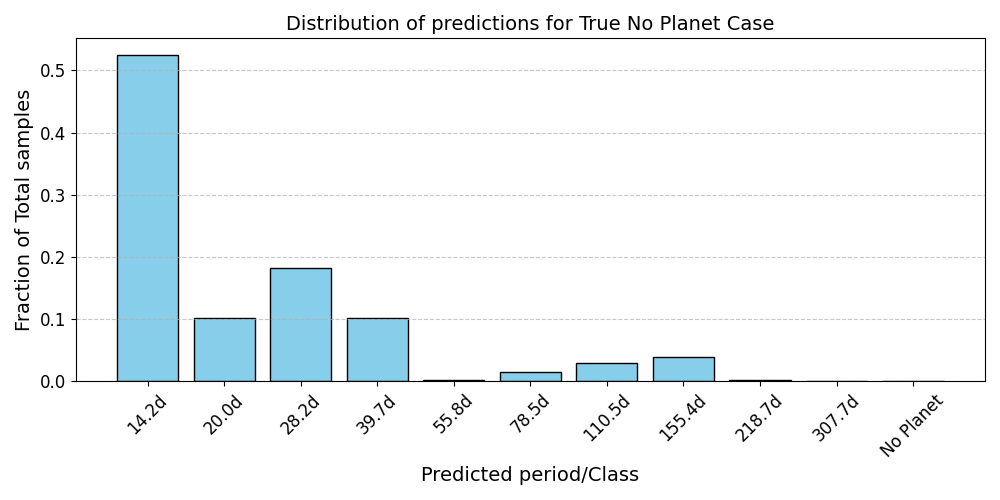}
    \caption{
    Distribution of predicted period classes for systems without planetary companions (the \enquote{No Planet} scenario). Period classes are shown in days. The distribution is heavily skewed toward the shortest period bin, with more than 50\% of the NP samples assigned to the lowest class despite the absence of a periodic signal. A smaller grouping is also visible near the solar rotation period (around 25–30 days), though it is far less prominent than the dominant low-period peak. This highlights the model’s tendency to predict short-period signals in the absence of true planetary signal, likely influenced by residual stellar variability or low-level noise mimicking short-timescale periodicity. This prediction pattern differs significantly from the typical orbital period distributions, reflecting the distinct nature of non-planetary light curves, characterized by residual stellar variability or noise rather than periodic transit-like features.
    }
    \label{fig:NPcase}
\end{figure*}

\begin{figure*}
    \center
    \includegraphics[scale=0.65]{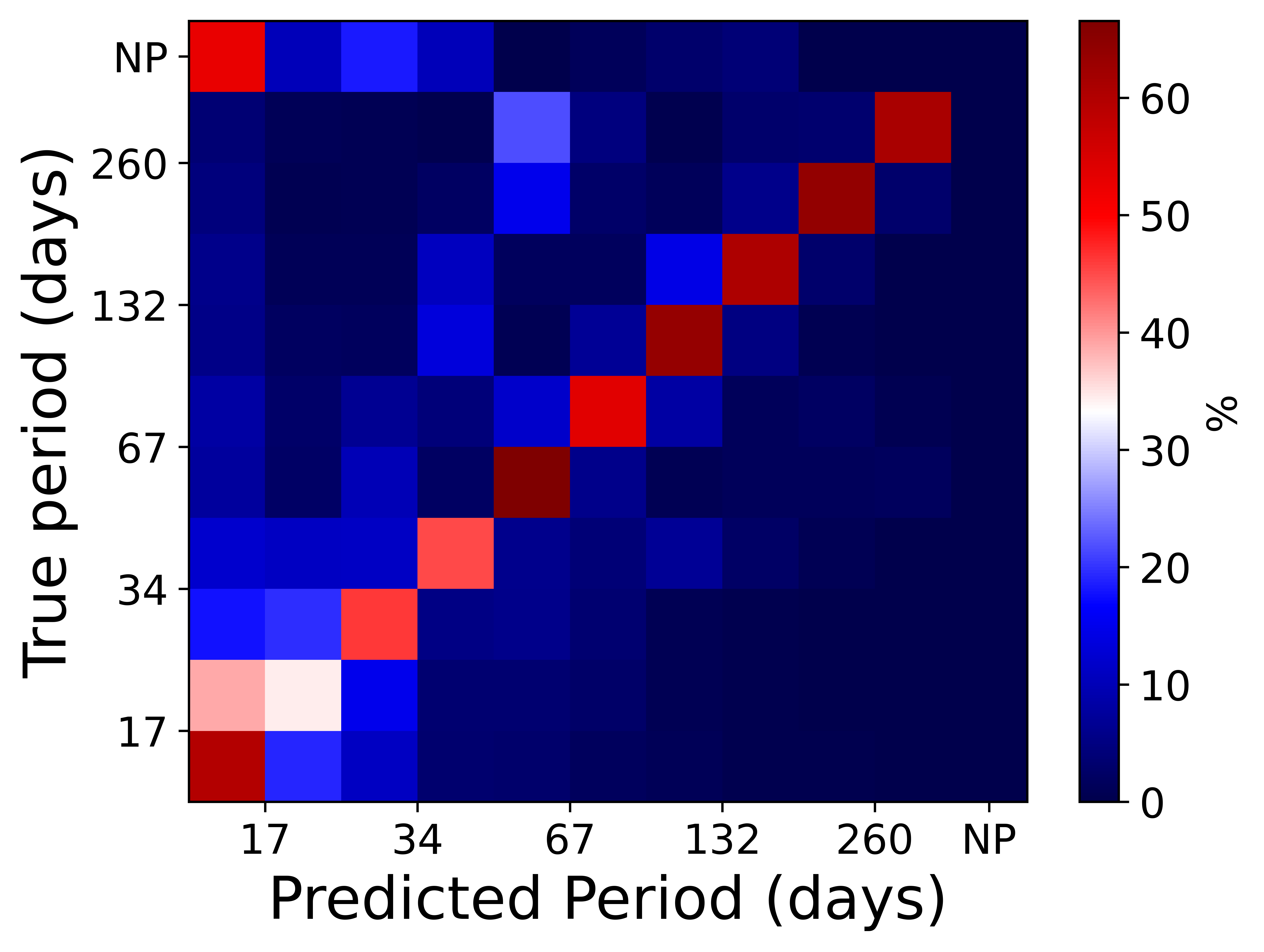}
    \caption{
    Normalized predicted probability distribution for all classes, including the \enquote{No Planet} (NP) scenario, from the ordered V3 dataset after fine-tuning. This figure displays data generated using the same procedure as in Figure~\ref{fig:my_lb17}, with the No Planet scenario explicitly shown for clarity. While the orbital period classes exhibit relatively sharp diagonal structure in the confusion matrix, the No Planet scenario shows a distinctly different distribution; heavily skewed toward the lowest period bin, as explained in Figure \ref{fig:NPcase}.
    }
    \label{fig:NPcfmatrix}
\end{figure*}

\section{The No Planet Scenario}

In Section~\ref{subsec:5.2.2}, we discussed model performance on realistic test data following fine-tuning, including challenges in identifying non-planetary systems. Here, we focus specifically on the \enquote{No Planet} scenario. Despite fine-tuning, the model frequently fails to recognize the absence of a planetary signal, producing spurious predictions for both period and semi-amplitude. To better understand this behavior, we examine the predicted distributions separately in the following subsections.

\label{appendix:Noplanetcase}

\subsection{Orbital Period Predictions}

The distribution of predicted periods for non-planetary systems is shown in Figure~\ref{fig:NPcase}. A substantial majority of these predictions ($\sim$91.2$\%$) fall below 45 days, with over half of the samples assigned to the lowest period bin. This indicates a strong bias of the model toward short-period predictions in the absence of a true Keplerian signal. A smaller secondary grouping is observed near the solar rotation period, though it is far less prominent than the dominant peak at the shortest bin.

\label{appendix:NoplanetcaseP}

While a few additional predictions appear at longer periods, they are relatively sparse and do not exhibit strong clustering. This behavior suggests that, rather than correctly identifying non-planetary systems, the model frequently defaults to spurious short-period solutions, potentially influenced by high-frequency stellar variability or noise. Consequently, when the model predicts periods greater than approximately 45 days, it is statistically more likely that the signal arises from a genuine planetary companion, as non-planetary classifications in this regime are rare. Even if the predicted period bin does not precisely match the true value, the underlying source of the detected periodic signal is still very likely planetary in nature (see Figure~\ref{fig:NPcfmatrix}).

\begin{figure*}
    \centering
    \includegraphics[scale = 0.85]{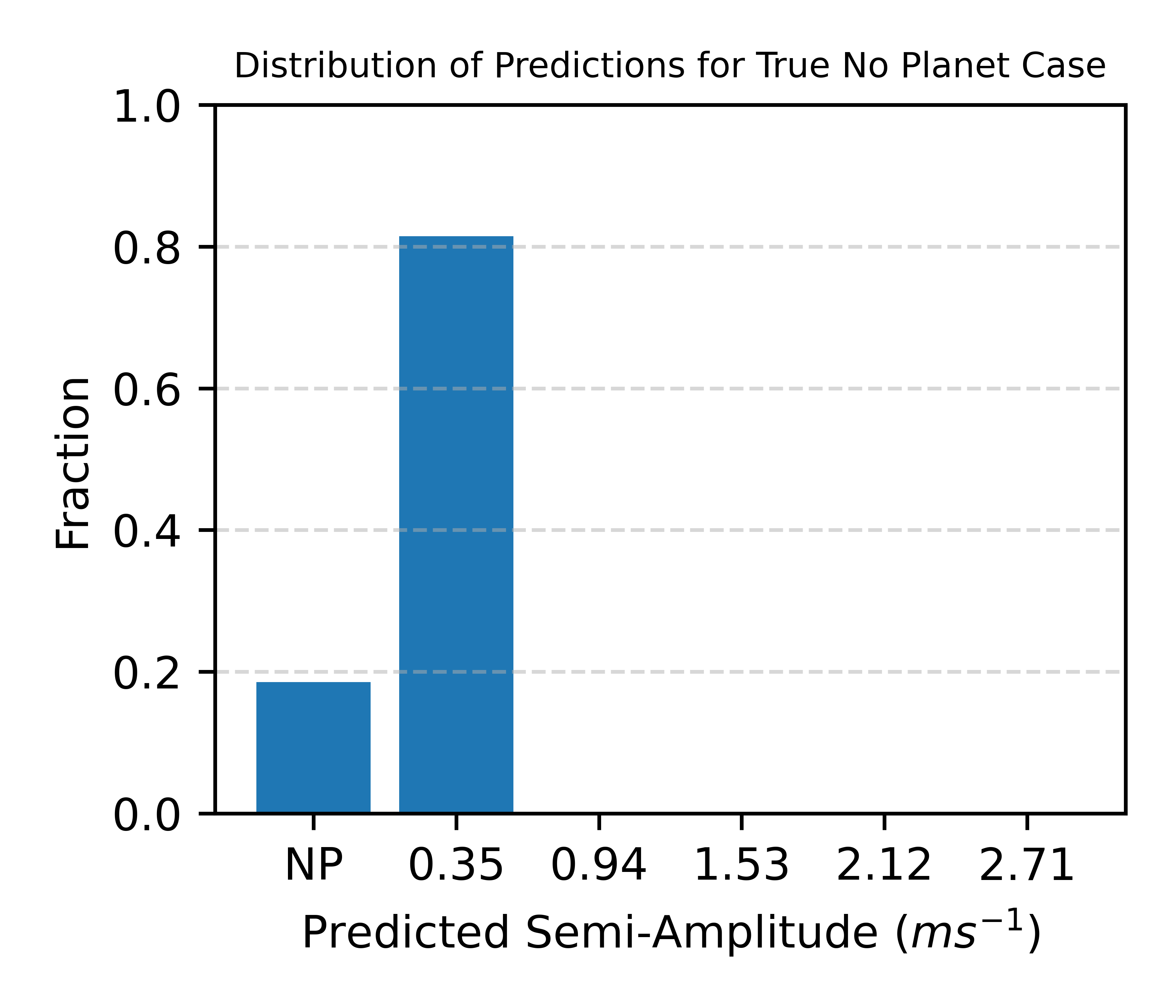}
    \caption{
    Distribution of predicted semi-amplitude classes for systems without injected planetary signal (the \enquote{No Planet} scenario).  Amplitudes are expressed in $ms^{-1}$. The predictions are heavily skewed toward the lowest amplitude bin, with over 80\% of samples assigned to this class. While the remaining predictions are primarily assigned to the correct No Planet class, the overall classification accuracy remains limited to approximately 20\%. This reflects the model’s tendency to infer low-amplitude planetary signals even when none are present. This highlights the model’s tendency to predict low amplitudes in the absence of a true planetary signal.
    }
    \label{fig:NPcaseK}
\end{figure*}

For completeness, we include the no-planet scenario within the confusion matrix shown in Figure~\ref{fig:my_label12}, and present the corresponding matrix separately in Figure~\ref{fig:NPcfmatrix} to highlight its distribution explicitly.

\subsection{Semi-Amplitude Predictions}

We previously examined semi-amplitude predictions for realistically ordered dataset samples in Section~\ref{subsec:5.2.3}, as illustrated in Figures~\ref{fig:amp_ftune} and~\ref{fig:amp_ftune_acc}. In that analysis, the \enquote{No Planet} scenario exhibited notably low classification accuracy. Figure~\ref{fig:NPcaseK} further explores this scenario by showing the distribution of predicted semi-amplitude classes for systems without planetary companions.

\label{appendix:NoplanetcaseK}

We find that approximately 80\% of the \enquote{No Planet} predictions fall into the lowest amplitude bin, while the remaining predictions are assigned to the correct \enquote{No Planet} class. Virtually no samples are misclassified into higher amplitude bins. This distribution suggests that a predicted amplitude class above the lowest bin is statistically unlikely to correspond to a non-planetary system. Consequently, such predictions may be interpreted as strong empirical indicators of a true planetary signal. However, further validation on real-world systems is required to substantiate this conclusion.

\bibliography{sample631}{}
\bibliographystyle{aasjournal}



\end{document}